\documentclass[twocolumn,showpacs,preprintnumbers,amsmath,amssymb]{revtex4}

\usepackage{graphicx}% include figure files
\usepackage{dcolumn}% Align table columns on decimal point
\usepackage{bm}% bold math
\usepackage{amsfonts}
\usepackage{amsmath}
\usepackage{amssymb}

\begin{document}

\title{\Large \bf Hyperdeformation in the cranked relativistic mean field 
theory: \\  the $Z=40-58$ part of nuclear chart}

\author{A.\ V.\ Afanasjev and  H.\ Abusara}

\address{Department of Physics and Astronomy, Mississippi State
University, MS 39762, USA} 

\date{\today}

\begin{abstract}
  The systematic investigation of hyperdeformation (HD) at high spin in the 
$Z=40-58$ part of the nuclear chart has been performed in the framework of 
the cranked relativistic mean field theory. The properties of the moments 
of inertia of the HD bands, the role of the single-particle and necking 
degrees of freedom at HD, the spins at which the HD bands become yrast, the 
possibility to observe discrete HD bands etc. are discussed in detail.
\end{abstract}


\maketitle

%%%%%%%%%%%%%%%%%%%%%%%%%%%%%%%%%%%%%%%%%%%%%%%%%%%%%%%%%%%%%%%%%%
\section{Introduction}
%%%%%%%%%%%%%%%%%%%%%%%%%%%%%%%%%%%%%%%%%%%%%%%%%%%%%%%%%%%%%%%%%%

    Since the discovery of superdeformation (SD) in $^{152}$Dy two decades
ago \cite{Dy152}, nuclear SD has been in the focus of attention of the nuclear 
structure community; it has been discovered in different mass regions and  
extensively studied experimentally \cite{SD-sys} and theoretically (see, 
for example, Refs.\ \cite{BHN.95,A150,Dudek} and references therein). New
phenomena such as identical bands \cite{BHN.95} were discovered, and rich 
variety of experimental data allowed to test modern theoretical tools under 
extreme conditions of large deformation and fast rotation.

   It was known for a long time from harmonic oscillator studies \cite{BM.75} 
that even more elongated shapes, called as hyperdeformed (HD) and characterized by 
the semi-axis ratio of around 3:1, are possible. The existence of such stable shapes
was later confirmed in the macroscopic+microscopic (MM) method 
\cite{BRAGSP.87,A180-HD-Chassman,Yb168,A.93,CNSPJ.94,WD.95,CR.95,JA.97,C.01}. 
Theoretical results on the states located in third (HD) minima are also available 
in self-consistent Hartree-Fock+Bogoliubov (HFB) approaches based on the Skyrme and 
Gogny forces (see Refs.\ \cite{ERC.97,SDN.06,HG.07} and references quoted therein), 
and relativistic mean field approach \cite{RMRG.95}. However, these results are 
restricted to spin zero states, which are difficult to measure in experiment. To 
our knowledge, the description of the HD states at high spin within the 
self-consistent approach has been attempted only in $^{108}$Cd \cite{AF.05-108Cd}
[within the cranked relativistic mean field (CRMF) method] and in four $A\sim 40$
mass nuclei \cite{IMYM.02} [within the cranked Skyrme-Hartree-Fock approach].
The general feature of all 
these calculations is the fact that the semi-axis ratio of the HD shapes is less 
than 3:1 \cite{Dudek}.

%------------------------------------------------------------------------------
\begin{figure*}[ht]
\includegraphics[angle=0,width=18cm]{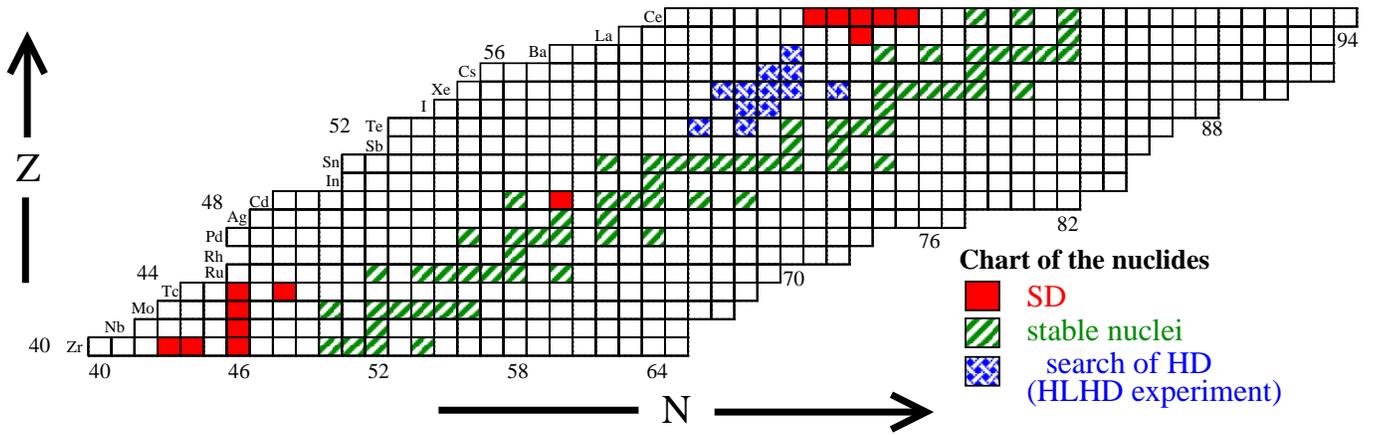}
\vspace{0.0cm}
\caption{(Color online) The chart of nuclei in the $Z=40-58$ region. Only experimentally 
known nuclei are shown. Experimental data on superdeformed nuclei are taken from Ref.\ 
\protect\cite{SD-sys}. The nuclei in which the search for HD structures has been 
performed in the HLHD experiment are taken from Ref.\ \protect\cite{Hetal.06}. 
}
\label{Nuclear-chart}
\end{figure*}
%------------------------------------------------------------------------------

 Let us mention two examples of such studies: one at spin zero, another at high spin.   
In actinide nuclei, the HD states 
are so-called third minima states around $^{232}$Th \cite{Th232,Th232-1,CNSPJ.94}. 
In these nuclei, the second saddle point is split, leading to the excited reflection-symmetric 
and reflection asymmetric configurations with large quadrupole and octupole deformations, 
$\beta_2 \sim 0.9$ and $\beta_3 \sim 0.35$. The density distribution at the HD minimum 
resembles a di-nucleus consisting of a nearly-spherical nucleus around the doubly-magic 
nucleus $^{132}$Sn and a well-deformed fragment from the neutron-rich $A\sim 100$ 
region \cite{CNSPJ.94}. Unfortunately, it is very difficult to study the HD 
states at low spin in experiment.  In order to overcome this problem,  
one should use the fact that the larger moment of inertia connected with the larger
deformation drives the nucleus towards larger deformations with increasing angular
momentum; the HD minimum is thus favored by rotation and becomes ultimately yrast at
high spin. For example, cranked Nilsson-Strutinsky calculations suggested the existence 
of very elongated high-spin minima in nuclei around  $^{168}$Yb \cite{Yb168}. These HD 
bands are expected to become yrast at spin around 80$\hbar$.

  On the experimental side, very little was known about hyperdeformation apart 
from some indications of this phenomenon at low spin in the uranium nuclei 
\cite{K.98} and light nuclei like $^{12}$C \cite{C12} and the observation
of the HD ridge structures at high spin in the $A\sim 150$ mass region 
\cite{152Dy-exp-1,152Dy-exp-2}. Recent observation of the very extended shapes in 
$^{108}$Cd \cite{Cd108-1,Cd108-2}, strongly motivated by earlier calculations of 
Ref.\ \cite{WD.95} and more recent studies of Ref.\ \cite{C.01}, has renewed interest 
in the study of hyperdeformation at high spin. Although the hyperdeformed nature
of the bands in this nucleus has not been confirmed in the subsequent cranked 
relativistic mean field analysis of Ref.\ \cite{AF.05-108Cd} (see also Sect.\ VB in
Ref.\ \cite{SDH.07}), this experiment provided a strong motivation for subsequent 
experimental searches in the $A\sim 125$ mass region (see Refs.\ 
\cite{HD-exp-1,HD-exp-2,Hetal.06}) and theoretical studies of Refs.\ \cite{Dudek,SDH.07} 
within the framework of the MM method. These experiments revealed rotational patterns 
in the form of ridge-structures in three-dimensional (3D) rotational mapped spectra with 
dynamic moments of inertia $J^{(2)}$ ranging from 63 to 111 MeV$^{-1}$ in 12 different 
nuclei \cite{Hetal.06}; the values around 110 MeV$^{-1}$ observed in $^{118}$Te, $^{124}$Xe 
and $^{124,125}$Cs suggest that the HD structures were populated in these experiments. 
However, no discrete rotational HD bands have been identified. It is also necessary to 
mention that several previous attempts to search for high spin HD structures in $^{147}$Gd 
\cite{147Gd-exp-1,147Gd-exp-2}, $^{152}$Dy \cite{152Dy-exp-1,152Dy-exp-2}, and $^{168}$Yb 
\cite{168Yb-exp-1} did not lead to convincing evidences for discrete  HD bands.

   So far, theoretical investigations of HD at high spin were carried out mainly in 
the framework of the MM method. One of the main goals of the current manuscript is to 
perform for the first time a systematic study of HD within the framework of fully 
self-consistent theory, the CRMF theory. Fig.\ \ref{Nuclear-chart} shows the part 
of the nuclear chart where our studies are performed. We restrict our investigation 
to even-even nuclei; the only exceptions are odd-mass nuclei $^{111}$I [in which
extremely SD doubly magic band has been found] and $^{123,124}$Xe, $^{123}$I and 
$^{125}$Cs [which are used in the study of the relative properties of the HD bands]. 
In each isotope chain we consider nuclei ranging from the most proton-rich 
ones up to the ones located at the neutron-rich side of the $\beta$-stability valley. 
Neutron-rich nuclei beyond the valley of the $\beta$-stability are excluded from 
consideration because of the experimental difficulties of studying them at high spins 
relevant for HD. With the goal to guide future experimental explorations and to find 
the nuclei in which the HD may be studied with current and future experimental facilities, 
we define the spins at which the HD bands become yrast in these nuclei. In addition, 
available experimental data on the HD ridge-structures in the Te, Xe, and Cs nuclei are 
analyzed. The general features of the HD bands are outlined.  

  The role of the single-particle degrees of freedom at hyperdeformation has not been 
studied in detail till now. One of the major goals of the current manuscript is the study 
of their role, and it is motivated by the desire to understand to what extent theoretical 
methods developed in the study of the SD bands are also applicable to the HD bands. It 
is very unlikely that the spins, parities and excitations energies of the HD bands will be 
known in the initial stage of their experimental study. The direct test of the structure 
of the wave functions of the single-nucleonic orbitals (e.g. via magnetic moments) will 
also not be possible at that stage. Thus, similar to  the case of superdeformation \cite{BRA.88,Rag.93,BHN.95,ALR.98}, 
the relative properties of different HD bands may play an important role in the interpretation 
of their structure. In this context, it is important to understand which changes of the 
single-particle orbitals are involved in going from one HD band to another, 
and how they affect physical observables like dynamic moments of inertia $J^{(2)}$, transition 
quadrupole moments $Q_t$, total spin $I$, etc. In particular, we will study whether the 
theoretical methods which were systematically used in the configuration assignment of the 
SD bands are also applicable to the HD bands. These include the methods based on the 
relative properties of the dynamic moments of inertia $J^{(2)}$ \cite{BRA.88,BHN.95}, on the 
effective alignments $i_{eff}$ \cite{Rag.93,BHN.95,ALR.98} and on the relative transition 
quadrupole moments $\Delta Q_t$ \cite{SDDN.96,MADLN.07}.

  The manuscript is organized as follows. 
The definition of physical observables  
and the details of numerical calculations are discussed in Sect.\ \ref{CRMF-theory}.
The spins at which the HD bands become yrast, the regions of nuclear chart where 
the experimental search for the HD structures may be successful and the general 
properties of the HD bands are outlined in Sect.\ \ref{HD-A120}. The data obtained 
in the search of the HD structures in the $A\sim 120$ mass region and the single-particle
degrees of freedom are also analysed in this section. Sec.\ \ref{I111-nucleus} is devoted 
to the analysis of extremely superdeformed (ESD) structure in $^{111}$I. The calculations 
predict the existence of doubly magic ESD structure in this nucleus with the deformations
 being close to HD, which may be observed with the current generation of $\gamma$-ray 
detectors. Finally, Sect.\ \ref{Concl-sect} contains the main conclusions of our work.

%%%%%%%%%%%%%%%%%%%%%%%%%%%%%%%%%%%%%%%%%%%%%%%%%%%%%%%%%%%%%%%%%%
\section{The details of the calculations}
\label{CRMF-theory}
%%%%%%%%%%%%%%%%%%%%%%%%%%%%%%%%%%%%%%%%%%%%%%%%%%%%%%%%%%%%%%%%%%

%%%%%%%%%%%%%%%%%%%%%%%%%%%%%%%%%%%%%%%%%%%%%%%%%%%%%%%%%%%%%%%%%%%
%\subsection{The CRMF Equations}
%%%%%%%%%%%%%%%%%%%%%%%%%%%%%%%%%%%%%%%%%%%%%%%%%%%%%%%%%%%%%%%%%%%%

  In the relativistic mean field (RMF) theory the nucleus is described 
as a system of point-like nucleons, Dirac spinors, coupled to mesons 
and to the photons \cite{SW.86,Reinh,VRAL.05}. The nucleons interact by the 
exchange of several mesons, namely a scalar meson $\sigma$ and three 
vector particles $\omega$, $\rho$ and the photon. The cranked relativistic 
mean field (CRMF) theory \cite{KR.89,KR.90,KR.93,A150}  represents the extension 
of RMF theory to the rotating frame. It has successfully been tested in a 
systematic way on the properties of different types of rotational bands in 
the regime of weak pairing such as normal-deformed \cite{AF.05}, 
superdeformed \cite{A60,A150} as well as smooth terminating bands 
\cite{VRAL.05}. 

  In the current study, we restrict ourselves to reflection 
symmetric shapes since previous calculations in the MM method show no 
indications that odd-multipole (octupole, ...) deformations play a role 
in the SD and HD bands of the nuclei covered by our study \cite{C.01} 
and in the HD bands of the $A\sim 110-125$ \cite{S.08} mass region.

%%%%%%%%%%%%%%%%%%%%%%%%%%%%%%%%%%%%%%%%%%%%%%%%%%%%%%%%%%%%%%%%%%%
\subsection{Physical observables}
%%%%%%%%%%%%%%%%%%%%%%%%%%%%%%%%%%%%%%%%%%%%%%%%%%%%%%%%%%%%%%%%%%%

 Similar to the case of the SD bands, it is reasonable to expect that the 
HD bands will not be linked to the low-spin level scheme for a long period 
of time. Thus, the spins and parities of the HD bands
will not be known and it will not be possible to define the kinematic 
moment of inertia $J^{(1)}$ since it depends on the absolute values of the 
spin. In such a situation, the dynamic moment of inertia $J^{(2)}$ will 
play an important role in our understanding of the structure of the
HD bands. This is similar to the case of the SD bands (see Refs.\ 
\cite{BRA.88,BHN.95}). Other observables, such as transition quadrupole moments 
$Q_t$ and effective (relative) alignments $i_{eff}$, will also be important.

  In the CRMF calculations, the rotational frequency ${\mathit\Omega}_x$, 
the kinematic moment of inertia $J^{(1)}$ and the dynamic moment of 
inertia $J^{(2)}$ are defined by 
\begin{eqnarray}
{\mathit\Omega}_x=\frac{dE}{dJ} ,
\label{Omega-def}
\end{eqnarray}
\begin{eqnarray}
J^{(1)}({\mathit\Omega}_x)&=&J\left\{\frac{dE}{dJ}\right\}^{-1} ,
\label{J1-def}\\
J^{(2)}({\mathit\Omega}_x)&=&\left\{\frac{d^2E}{dJ^2}\right\}^{-1} .
\label{J2-def}
\end{eqnarray}

 The charge  quadrupole $Q_0$ and mass hexadecupole $Q_{40}$ 
moments are calculated by using the expressions
\begin{eqnarray}
Q_0&=&e\sqrt{\frac{16\pi}{5}}
\sqrt{\left\langle r^2Y_{20}\right\rangle_p^2
+2\left\langle r^2Y_{22}\right\rangle_p^2} ,
\label{Q_0-def} \\
Q_{40}&=&\left\langle r^4Y_{40}\right\rangle _p+\left\langle
r^4Y_{40}\right\rangle _n ,
\label{Q_40-def}
\end{eqnarray}
where the labels $p$ and $n$ are used for protons and neutrons, respectively, 
and $e$ is the electrical charge. At axially symmetric shapes, typical for 
the hyperdeformed states, the transition quadrupole moment $Q_t$ is equal 
to $Q_0$.

   The quadrupole deformation $\beta_2$ for axially-symmetric shapes is 
frequently defined in self-consistent calculations from calculated and/or 
experimental quadrupole moments using simple relation 
\cite{HG.07,A250,SGP.05}
\begin{eqnarray} 
\beta_2 = \frac{1}{XR^2}\sqrt{\frac{5\pi}{9}} Q_{0}^X ,
\label{beta2}
\end{eqnarray}
where $R=1.2 A^{1/3}$ fm is the radius of the nucleus, and $Q_{0}^X$ is a quadrupole 
moment of the $X$-th (sub)system expressed in fm$^2$. Here $X$ refers either to proton 
($X=Z$) or neutron ($X=N$) subsystem or represents total nuclear system ($X=A$). 
This expression, however, neglects the higher powers of $\beta_2$ and higher 
multipolarity deformations $\beta_4, \beta_6, ...$  \cite{NR.96}, which play 
an important role at hyperdeformation.

  Considering that the definition of the deformation is model dependent \cite{NR.96}, 
and that this quantity is not experimentally measurable, we prefer to use transition 
quadrupole moment $Q_t$ for the description of deformation properties of hyperdeformed 
states. This is experimentally measurable quantity, so in the future our 
predictions can be directly compared with experiment. The deformation properties of 
the yrast SD band in $^{152}$Dy (which is one of the most deformed SD bands \cite{Cd108-1}) 
are used as a reference. This is done by introducing normalized transition quadrupole 
moment $Q_t^{norm}(Z,A)$ in the $(Z,A)$ system
\begin{eqnarray}
Q_t^{norm}(Z,A)=\frac{ZA^{2/3}}{100.36}\,\,\, e{\rm b}
\label{norm-Q}
\end{eqnarray} 
This equation is based on the ratio $Q_t^{norm}(Z,A)/Q_t(^{\rm 152}{\rm Dy})$ calculated 
using  Eq.\ (\ref{beta2}) under the assumption that the $\beta_2$-values in the $(Z,A)$ 
system and in $^{152}$Dy are the same. We use the value $Q_t(^{152}{\rm Dy})=18.73$ 
$e$b obtained in the CRMF calculations with the NL1 parametrization of the RMF
Lagrangian for the yrast SD band in $^{152}$Dy  at $I=60\hbar$ in Ref.\ \cite{A150}. 
Thus, in first approximation (neglecting the higher powers of $\beta_2$ and higher multipolarity deformations $\beta_4, \beta_6, ...)$) the equilibrium deformation of the 
band  in the $(Z,A)$ system having the $Q_t^{norm}(Z,A)$ value is the same as in the 
yrast SD band of $^{152}$Dy. We describe the band as hyperdeformed if its $Q_t$ value 
exceeds $Q_t^{norm}(Z,A)$ by at least 40\%. This criteria is somewhat relaxed in the 
$Z=40,\,\,42,\,\,44$ nuclei for which the band is defined as HD if its $Q_t$ value 
exceeds $Q_t^{norm}(Z,A)$ by at least 30\%.

  The effective (relative) alignment $i_{eff}$ between two bands is defined as the difference 
between the spins of two levels in bands A and B at the same rotational frequency 
$\Omega_x$ \cite{Rag.93}:
\begin{eqnarray}
i_{eff}^{B,A}(\Omega_x)=I_B(\Omega_x)-I_A(\Omega_x)
\end{eqnarray}
This quantity has been used frequently in the analysis of the single-particle
structure of the SD bands and the configuration assignment (see Refs.\ 
\cite{Rag.93,ALR.98} and references quoted therein). It depends on both the alignment 
properties of the single-particle orbitals(s) by  which the two bands differ and  
the polarization effects induced by the particles in these orbitals \cite{AR.00}. The 
latter are in part related to nuclear magnetism.

 Because the pairing correlations are relatively weak in the HD bands of 
interest (see Sect.\ \ref{Xe124-neigh}), their intrinsic structure can be 
described by means of the dominant single-particle components of the 
hyperintruder states occupied. The calculated configurations will be 
labeled by $[p,n_1n_2]$, where $p$, $n_1$ and $n_2$ are the number of 
proton $N=7$ and neutron $N=7$ and $N=8$ hyperintruder orbitals occupied, 
respectively.  For most of the HD configurations, neutron $N=8$ orbitals 
are not occupied, so the label ${n_2}$  will be omitted in the labeling 
of such configurations.

Single-particle orbitals are labeled by $[Nn_z\Lambda]\Omega^{sign}$.
$[Nn_z\Lambda]\Omega$ are the asymptotic quantum numbers (Nilsson quantum numbers) 
of the dominant component of the wave function at $\Omega_x=0.0$ MeV. 
The superscripts {\it sign} to the orbital labels are used to indicate the sign 
of the signature $r$ for that orbital $(r=\pm i)$.

   The spins at which the SD and HD configurations become yrast in the 
calculations are defined as crossing spins $I_{cr}^{SD}$ and 
$I_{cr}^{HD}$, respectively.

%%%%%%%%%%%%%%%%%%%%%%%%%%%%%%%%%%%%%%%%%%%%%%%%%%%%%%%%%%%%%%%%%%%%%%%%%%%%%%%%%%%
\begin{figure}[ht]
\vspace{0.7cm}
\includegraphics[angle=0,width=7.5cm]{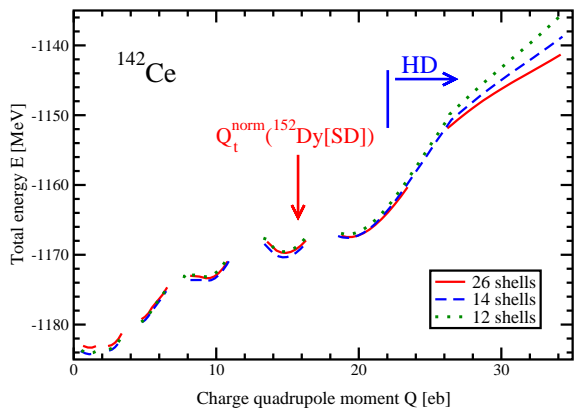}
\vspace{0.2cm}
\caption{ (Color online) Potential energy surfaces (PES) obtained in the axially 
symmetric RMF calculations without pairing in the $^{142}$Ce nucleus. The results 
of calculations with $N_F=12$, 14 and 26 are shown. In all these calculations, $N_B$ 
is fixed at 26. The results with $N_F=26$ correspond to a fully converged solution: 
the binding energies do not change with further increase of $N_F$. The gaps in the 
PES lines are due to jumps of the solution from one single-particle configuration 
to another. The same single-particle configurations are compared at the same value 
of charge quadrupole moment. The normalized value of transition quadrupole moment 
$Q^{norm}_t$ corresponding to the deformation of the yrast SD band in $^{152}$Dy is 
indicated by arrow. The range of hyperdeformation is also indicated.}
\label{142Ce-pes}
\end{figure}
%%%%%%%%%%%%%%%%%%%%%%%%%%%%%%%%%%%%%%%%%%%%%%%%%%%%%%%%%%%%%%%%%%%%%%%%%%%%%%%%%%%

%%%%%%%%%%%%%%%%%%%%%%%%%%%%%%%%%%%%%%%%%%%%%%%%%%%%%%
\subsection{Numerical scheme of the CRMF calculations}
%%%%%%%%%%%%%%%%%%%%%%%%%%%%%%%%%%%%%%%%%%%%%%%%%%%%%%

  The CRMF equations are solved in the basis of an anisotropic three-dimensional 
harmonic oscillator in Cartesian coordinates characterized by the deformation 
parameters $\beta_0$ and $\gamma$ and oscillator frequency $\hbar \omega_0= 41 
A^{-1/3}$ MeV, for details see Refs. \cite{KR.89,A150}. The truncation of basis is
performed in such a way that all states belonging to the shells up to fermionic 
$N_F$ and bosonic $N_B$ are taken into account.

  The impact of the truncation of basis on the numerical accuracy of
the calculations has first been studied in the axially symmetric RMF code, 
see Fig.\ \ref{142Ce-pes}.  In the mass region of interest, the calculations with $N_F=12$ 
provide a reasonable approximation to the fully convergent $N_F=26$ solution up to 
a deformation typical for the SD shapes. However, this truncation scheme becomes a 
poor approximation when the quadrupole moment appreciably exceeds the one corresponding 
to the lower limit of HD; the difference between the $N_F=12$ and $N_F=26$ solutions 
increases rapidly with the increase of quadrupole moment (see Fig.\ \ref{142Ce-pes}). 
On the other hand, in this quadrupole moment range the results of the calculations 
with $N_F=14$ are closer to exact solution, although still exceeding it by $\sim 1-2$ 
MeV at the upper end of the calculated quadrupole moment range. It was tested that 
with the decrease of the mass, the difference between the $N_F=14$ and $N_F=26$ solutions 
will also decrease as well, so that the difference falls within the range of 1 MeV for 
the majority of the nuclei under study.

%%%%%%%%%%%%%%%%%%%%%%%%%%%%%%%%%%%%%%%%%%%%%%%%%%%%%%%%%%%%%%%%%%%%%%%%%%%%%%%%%%%
\begin{figure}[ht]
\includegraphics[angle=0,width=6.7cm]{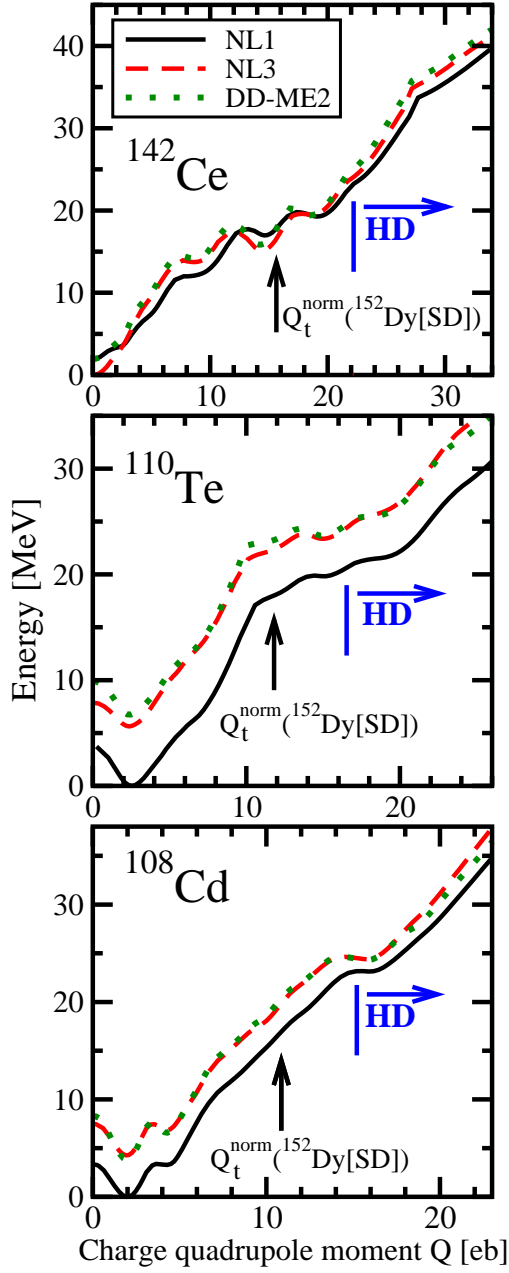}
\vspace{0.2cm}
\caption{(Color online) The same as in Fig.\ \ref{142Ce-pes}, but for the 
results obtained in the axially symmetric RMF calculations with 
pairing using different parametrizations of the RMF Lagrangian
and $N_F=26$. Figure shows the binding energies normalized with 
respect to the lowest energy of the lowest potential energy 
curve.}
\label{PES-diff-par}
\end{figure}
%%%%%%%%%%%%%%%%%%%%%%%%%%%%%%%%%%%%%%%%%%%%%%%%%%%%%%%%%%%%%%%%%%%%%%%%%%%%%%%%%%%

%----------------------------------------------------------------------------
\begin{figure}[h]
\vspace{1.0cm}
\includegraphics[angle=0,width=7.5cm]{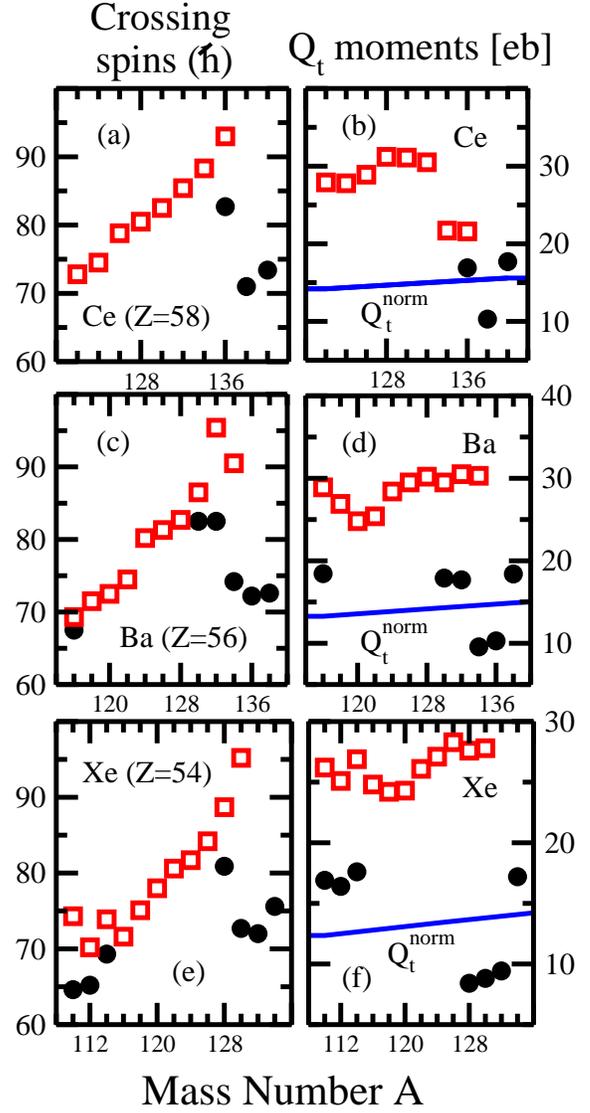}
\vspace{0.45cm}
\caption{(Color online) The crossing spins (left panels) at which the SD 
(solid circles) and HD (open squares) configurations become yrast and 
their transition quadrupole moments $Q_t$ (right panels) for the Ce, 
Ba and Xe isotopes. The values for the SD configurations are shown only 
in the cases when they become yrast at lower spins than the HD configurations. 
The normalized transition quadrupole moments $Q^{norm}_t$ corresponding 
to the deformation of the yrast SD band in $^{152}$Dy are also shown.}
\label{Icross-Ce-Ba-Xe}
\end{figure}
%---------------------------------------------------------------------------

 These conclusions have also been tested in triaxial CRMF calculations. 
It was concluded that physical observables of interest 
are described with sufficient numerical accuracy when $N_F=12$ is used for the 
SD and ND states and $N_F=14$ for the HD states. 
 Thus, we employ a hybrid calculational scheme 
in which the CRMF solutions in the ND- and SD minima are sought using $N_F=12$, while 
the ones in the HD minima using $N_F=14$. In all CRMF calculations, we use $N_B=20$.
In order to eliminate the numerical 
inaccuracies in the definition of the crossing spin $I_{cr}^{HD}$, the yrast ND/SD 
configurations, which are crossed by the yrast HD configuration, were recalculated 
in the crossing region using $N_F=14$, and only then the crossing spin was defined. 
One should keep in mind that even with $N_F=14$ the spins at which the HD configurations 
become yrast in the calculations may be overestimated by $1-2\hbar$ when the deformation 
of the HD configurations exceeds appreciably the one corresponding 
to the lower limit of HD.

  When searching for different types of rotational structures it is important to find
the solutions in all local minima which are close to the yrast line in order to properly define 
the crossing spins between the rotational structures of different nature.
This is easily achievable in the macroscopic+microscopic approach by creating 
potential energy surfaces (PES) in the deformation space covering quadrupole and 
triaxial deformations \cite{WD.95,PhysRep}. However, the computational cost to
create similar PES in the self-consistent models is enormous, thus, it has never 
been attempted in rotating nuclei. In order to overcome this problem, we use 
the fact that in self-consistent approaches without pairing the deformation of the 
basis defines to a large extent the local minima where the solutions will be obtained. 
Thus, the solutions in the ND minima, including triaxial ones, are searched using three 
combinations of the deformation of basis: $(\beta_0=0.30, \gamma=-30^{\circ})$, 
$(\beta_0=0.30, \gamma=0^{\circ})$, and $(\beta_0=0.30, \gamma=+30^{\circ})$. In a 
similar way, the solutions in the SD minima are searched using the following 
combinations of the deformations of basis $(\beta_0=0.65, \gamma=-30^{\circ})$, 
$(\beta_0=0.65, \gamma=0^{\circ})$, $(\beta_0=0.65, \gamma=+30^{\circ})$, and 
$(\beta_0=0.8, \gamma=0^{\circ})$. The latter deformation of basis also leads frequently 
to the HD solutions. The deformation of basis $(\beta_0=1.0, \gamma=0^{\circ})$ 
has been used for the search of the solutions in the HD minima. Non-zero 
$\gamma$-deformations of basis at large $\beta_0$ lead either to the same solution 
as $\gamma=0^{\circ}$ or to the highly excited configurations. 
For each of the above mentioned values of the deformation of basis, the lowest in energy 
solutions are calculated as a function of spin, and the yrast line is formed from 
these solutions.

%----------------------------------------------------------------------------
\begin{figure}[h]
\vspace{1.0cm}
\includegraphics[angle=0,width=6.5cm]{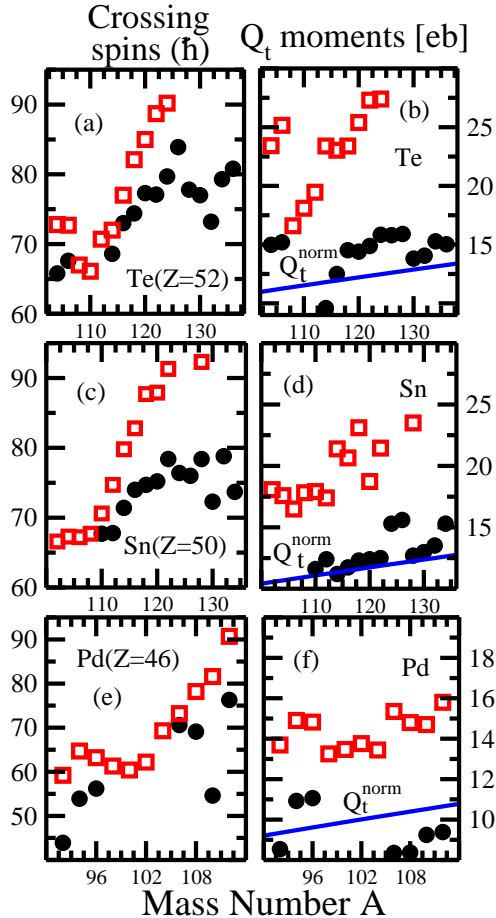}
\vspace{0.45cm}
\caption{(Color online) The same as in Fig.\ \ref{Icross-Ce-Ba-Xe}, but for 
the Te, Sn, and Pd isotopes.}
\label{Icross-Te-Sn-Pd}
\end{figure}
%---------------------------------------------------------------------------

%%%%%%%%%%%%%%%%%%%%%%%%%%%%%%%%%%%%%%%%%%%%%%%%%%%%%%%%%%%%%%%%%%%%
\subsection{The selection of the RMF parametrization.}
%%%%%%%%%%%%%%%%%%%%%%%%%%%%%%%%%%%%%%%%%%%%%%%%%%%%%%%%%%%%%%%%%%%%

   The NL1 parametrization of the RMF Lagrangian \cite{NL1} is used in the
majority of the calculations in the current manuscript. As follows from previous 
studies, this parametrization provides a good description of the moments of inertia 
of the rotational bands in unpaired regime in the SD and ND minima 
\cite{A150,ALR.98,A60,VRAL.05}, the single-particle energies for the nuclei around 
the valley of $\beta$ stability \cite{ALR.98,A250} and the excitation energies of the 
SD minima \cite{LR.98}. NL3 \cite{NL3} is an alternative parametrization, the quality 
of which has been tested in rotating nuclei (but less extensively than in the case 
of NL1) \cite{ALR.98,Zn68,A60,AF.05}. Some results with this parametrization 
will be presented. Few results obtained with the NLSH \cite{NLSH} and NLZ 
\cite{NLZ} parametrizations will be shown in Sect.\ \ref{Xe124-neigh}
in order to illustrate the possible spread of calculated  quantities. It is necessary 
to keep in mind that the quality of the NLSH parametrization in respect of the 
description of rotational properties of the nuclei as well as their single-particle 
energies is not as good as that of the NL1 and NL3 \cite{A60,ALR.98,A250}, and 
the force NLZ has not been tested in that respect.

   The spins at which the rotational structures belonging to different minima 
in potential energy surfaces become yrast depend in general on the relative 
energies of these minima and on the moments of inertia of rotational structures in 
these minima. Previous experience shows that different parametrizations of the RMF 
Lagrangian give similar moments of inertia for the same configuration 
\cite{ALR.98,A60,Zn68,VRAL.05} (see also Fig.\ \ref{Xe124-j1-j2} below). Fig.\ \ref{PES-diff-par} 
also illustrates that the 
potential energy surfaces at spin zero as a function of charge quadrupole moment obtained 
with the NL1 and NL3 parametrizations are similar in shape. These two facts suggest 
that the HD configurations should become yrast at approximately the same spins in 
both parametrizations: this conclusion is confirmed in Sect.\ \ref{sys-Icros}. It is 
interesting to note that the NL3 curve in Fig.\ \ref{PES-diff-par}  is similar to 
the one obtained with recently developed density-dependent meson-exchange effective 
interaction DD-ME2 \cite{LNVR.05}, which represents a new class of the RMF 
parametrizations as compared with NL1 and NL3. However, so far this interaction has not 
been used in the studies of rotating nuclei, thus, it is not employed in 
the current study since its reliability in the description of rotational properties 
is not known.

%----------------------------------------------------------------------------
\begin{figure}[ht]
\vspace{1.0cm}
\includegraphics[angle=0,width=8cm]{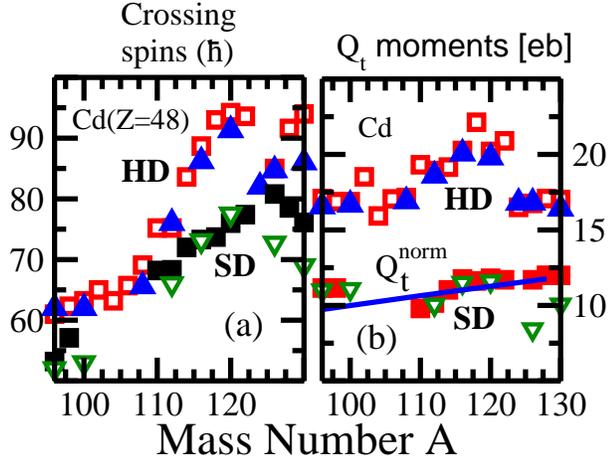}
\caption{(Color online) The same as in Fig.\ \ref{Icross-Ce-Ba-Xe}, but for 
the Cd isotopes. The results of the calculations with the NL1 (HD - open 
squares, SD - solid squares) and NL3 (HD - solid triangles up, SD - open 
triangles down) parametrizations of the RMF Lagrangian are presented. 
Note that the calculations with NL3 were performed only for selected 
nuclei.}
\label{Icross-Cd}
\end{figure}
%---------------------------------------------------------------------------

%%%%%%%%%%%%%%%%%%%%%%%%%%%%%%%%%%%%%%%%%%%%%%%%%%%%%%%%%%%%%%%%%%%%%%
\section{Hyperdeformation at high spin: where to expect and its general 
features}
\label{HD-A120}
%%%%%%%%%%%%%%%%%%%%%%%%%%%%%%%%%%%%%%%%%%%%%%%%%%%%%%%%%%%%%%%%%%%%%%

%%%%%%%%%%%%%%%%%%%%%%%%%%%%%%%%%%%%%%%%%%%%%%%%%%%%%%%%%%%%%%%%%%%%%%%%%
\subsection{The systematics of crossing spins and transition quadrupole
moments of the HD bands}
\label{sys-Icros}
%%%%%%%%%%%%%%%%%%%%%%%%%%%%%%%%%%%%%%%%%%%%%%%%%%%%%%%%%%%%%%%%%%%%%%%%%

\begin{figure}[h]
\vspace{1.0cm}
\includegraphics[angle=0,width=6.5cm]{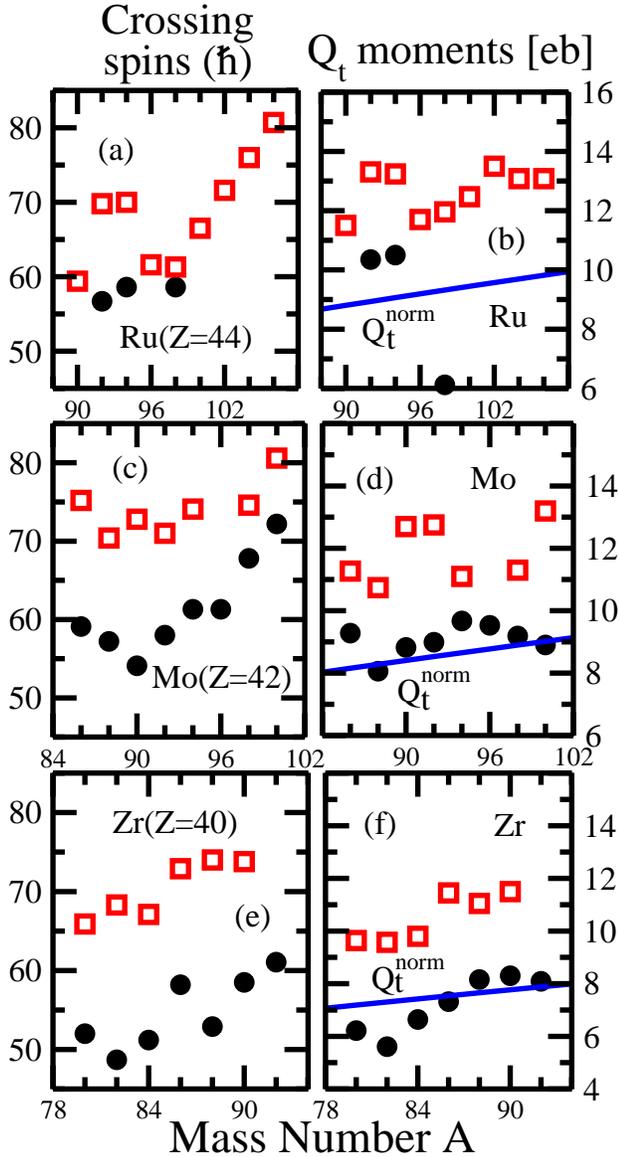}
\vspace{0.45cm}
\caption{(Color online) The same as in Fig.\ \ref{Icross-Ce-Ba-Xe}, but for 
the Ru, Mo and Zr isotopes.}
\label{Icross-44-42-Zr}
\end{figure}

 Figs.\ \ref{Icross-Ce-Ba-Xe}, \ref{Icross-Te-Sn-Pd}, \ref{Icross-Cd} 
and \ref{Icross-44-42-Zr} display the spins at which the SD and HD 
configurations become yrast (crossing spins) in the CRMF calculations.  
In addition, the calculated transition quadrupole moments of these 
configurations at spin values close to the crossing spins are shown. 
The calculated HD configurations are near-prolate. One can see that the 
crossing spins $I_{cr}^{HD}$ are typically lower for proton-rich nuclei. 
Such a feature is seen in most of the isotope chains; by going from the 
$\beta$-stability valley toward the proton-drip line, one can lower 
$I_{cr}^{HD}$ by approximately $10\hbar$. The minimum of crossing spins 
$I^{HD}_{cr}$ is reached at $N\approx Z+10$ in the Pd, Te and Ru 
isotope chains (see Figs.\  \ref{Icross-Te-Sn-Pd}e, \ref{Icross-Te-Sn-Pd}a  
and \ref{Icross-44-42-Zr}a), and the Mo isotope chain (Fig.\ \ref{Icross-44-42-Zr}c) 
shows almost no dependence of  $I^{HD}_{cr}$ on mass number. In other isotope 
chains, the minima in crossing spins $I^{HD}_{cr}$ appear in most proton-rich nuclei. 
Considering that the sensitivity of modern $\gamma$-ray detectors 
allows to study discrete rotational bands only up to $\approx 65\hbar$ 
in medium mass nuclei \cite{Dy156,152Dy-link,Ce132-131}, and that the observation 
of higher spin states will most likely require a new generation of 
$\gamma$-ray tracking detectors such as GRETA or AGATA, these features 
of crossing spins $I^{HD}_{cr}$ represent an important constraint.

  As suggested by the studies of the Jacobi shape transition in  Ref.\ 
\cite{SDH.07},  the coexistence of the SD and HD minima at the feeding 
spins may have an impact on the survival of the HD minima because of the 
decay from the HD to  SD configurations. If this mechanism is active, 
then only the nuclei in which the HD minimum is lower in energy than the 
SD one at the feeding spin and/or the nuclei characterized by the large 
barrier between the HD and SD minima will be the reasonable candidates 
for a search of the HD bands. Figs.\ \ref{Icross-Ce-Ba-Xe}, \ref{Icross-Te-Sn-Pd}, 
\ref{Icross-Cd} and \ref{Icross-44-42-Zr} show that the HD configurations become 
yrast at lower spin than the SD ones only in a specific mass range which 
depends on the isotope chain. This range can be narrow as in the case of Te 
isotopes (Fig.\ \ref{Icross-Te-Sn-Pd}a) or wide as in the case of Ce isotopes 
(Fig.\ \ref{Icross-Ce-Ba-Xe}a). The question of the population of the HD bands 
within the RMF framework definitely deserves an additional study, but such a 
study is beyond the scope of the present manuscript.

   Fig.\ \ref{Icross-Cd} compares the results of the calculations for Cd isotopes 
obtained with the NL1 and NL3 parametrizations of the RMF Lagrangian. One can see 
that both parametrizations predict similar crossing spins $I^{SD}_{cr}$ and 
$I^{HD}_{cr}$ and similar transition quadrupole moments. However, in average,
the crossing spins $I^{HD}_{cr}$ calculated with NL3 are somewhat lower (by 
$1-2\hbar$) than the ones obtained in the calculations with NL1.

%%%%%%%%%%%%%%%%%%%%%%%%%%%%%%%%%%%%%%%%%%%%%%%%%%%%%%%%%%%%%%%%%%%%%
\subsection{The $A\sim 120$ region: the analysis of experimental data}
%%%%%%%%%%%%%%%%%%%%%%%%%%%%%%%%%%%%%%%%%%%%%%%%%%%%%%%%%%%%%%%%%%%%%

%-----------------------------------------------------------------------------
\begin{table}[h]
\caption{The values of the dynamic moment of inertia $J^{(2)}_{exp}$ 
of ridge structures measured in the HLHD experiment \cite{Hetal.06}. 
Theoretical results obtained in the MM calculations \cite{SDH.07} are 
shown in the last column.}
\newcommand{\m}{\hphantom{$-$}}
\newcommand{\cc}[1]{\multicolumn{1}{c}{#1}}
\renewcommand{\tabcolsep}{0.5pc} % enlarge column spacing
\renewcommand{\arraystretch}{1.4} % enlarge line spacing
\begin{tabular}{@{}lcc}
\hline
Nucleus    & $J^{(2)}_{exp}$   & $J^{(2)}_{MM}$ \\ \hline$^{126}$Ba  & 77                        &      118   \\                
$^{123}$Xe  & 71                        &            \\                
$^{122}$Xe  & 77                        &      108   \\              
$^{121}$Xe  & 63                        &            \\              
$^{120}$Te  & 71                        &            \\              
$^{118}$Te  & 111                       &      97    \\              
$^{125}$Cs  & 100                       &      106   \\              
$^{124}$Cs  & 111                       &            \\         
$^{124}$Xe  & 111                       &      111   \\              
$^{122}$I   & 71                        &            \\              
$^{121}$I   & 77                        &      102   \\              
$^{126}$Xe  & 83                        &      110   \\ \hline
\end{tabular}\\[2pt]
\label{HD-table}
\end{table}
%-----------------------------------------------------------------------------
%----------------------------------------------------------------------------
%---------------------------------------------------------------------------

  Recent Hyper-Long-HyperDeformed (HLHD) experiment at the 
EUROBALL-IV $\gamma$-detector array revealed some features expected 
for HD nuclei \cite{HD-exp-1,HD-exp-2,Hetal.06}. Although no discrete 
HD rotational bands have been identified, rotational 
patterns in the form of ridge-structures in three-dimensional (3D) rotational 
mapped spectra are identified with dynamic moments of inertia $J^{(2)}$ ranging 
from 71 to 111 MeV$^{-1}$ in 12 different nuclei selected by charged particle- 
and/or $\gamma$-gating (see Table \ref{HD-table}). The four nuclei, $^{118}$Te, 
$^{124}$Cs, $^{125}$Cs and $^{124}$Xe, found with moment of inertia $J^{(2)}\sim 110$
MeV$^{-1}$ are most likely hyperdeformed \footnote{The HD ridges in $^{152}$Dy
are characterized by $J^{(2)}\sim 130$ MeV$^{-1}$ \cite{152Dy-exp-2}.} while 
the remaining nuclei with smaller values of $J^{(2)}$ are expected to be 
superdeformed. The width in energy of the observed ridges indicates that there 
are $\approx 6-10$ transitions in the HD cascades, and a fluctuation analysis 
shows that the number of bands in the ridges exceeds 10. The HD ridges are
observed in the frequency range of about 650 to 800 keV, and their dynamic 
moments of inertia have typical uncertainty of 10\% (e.g. $111\pm 11$ MeV$^{-1}$ 
in $^{124}$Xe) \cite{Hub-private}.

%%%%%%%%%%%%%%%%%%%%%%%%%%%%%%%%%%%%%%%%%%%%%%%%%%%%%%%%%%%%%%%%%%%%%%%%%%%%%%%%%%
\begin{figure}[ht]
\vspace{0.8cm}
\includegraphics[angle=0,width=8.0cm]{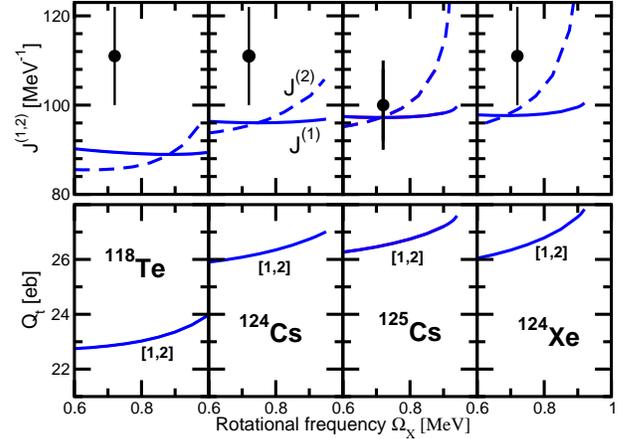}
\caption{(Color online) Calculated kinematic and dynamic moments of inertia 
(top panels) and transition quadrupole moments (bottom panels) as a function 
of rotational frequency for the lowest HD solutions in $^{118}$Te, 
$^{124,125}$Cs and $^{124}$Xe. The structure of calculated configurations is 
indicated at bottom panels. Experimental data for dynamic moments of inertia 
of ridge structures are shown in top panels.}
\vspace{0.0cm}
\label{HD-exp}
\end{figure}
%%%%%%%%%%%%%%%%%%%%%%%%%%%%%%%%%%%%%%%%%%%%%%%%%%%%%%%%%%%%%%%%%%%%%%%%%%%%%%%%%%

%%%%%%%%%%%%%%%%%%%%%%%%%%%%%%%%%%%%%%%%%%%%%%%%%%%%%%%%%%%%%%%%%%%%%%%%%%%%%%%%%%%
\begin{figure}[ht]
\vspace{0.7cm}
\includegraphics[angle=0,width=8.0cm]{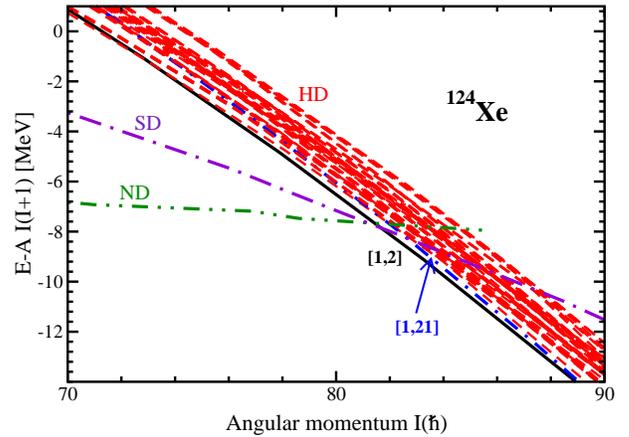}
\caption{ (Color online) Energies of the calculated configurations relative to a 
smooth liquid drop reference $AI(I+1)$, with the inertia parameter $A=0.01$. The 
ND and SD yrast lines are shown by dotted and dot-dot-dashed lines, respectively. 
Solid and dot-dashed lines are used for the [1,2] and [1,21] HD configurations, 
respectively. Dashed lines represent excited HD configurations.}
\label{Xe124-eld-NL1}
\end{figure}
%%%%%%%%%%%%%%%%%%%%%%%%%%%%%%%%%%%%%%%%%%%%%%%%%%%%%%%%%%%%%%%%%%%%%%%%%%%%%%%%%%%

 The experimental data show unusual features never before seen in the studies 
of the SD bands.  For example, the addition of one neutron on going from 
$^{124}$Cs to $^{125}$Cs decreases the experimental $J^{(2)}$ value by $\sim 10\%$ 
(from 111 MeV$^{-1}$ down to 100 MeV$^{-1}$, see Table \ref{HD-table}). A similar 
situation is also seen in the SD minimum: the addition of one neutron on going 
from $^{121}$Xe to $^{122}$Xe increases the experimental $J^{(2)}$ value by 
$\sim 22\%$ (from 63 MeV$^{-1}$ to 77 MeV$^{-1}$, see Table \ref{HD-table}). It 
is impossible to find an explanation for such a big impact of the single particle 
on the properties of nuclei: previous studies in the SD minima in different parts 
of the nuclear chart never showed such features. The case of the pair of $^{123}$Xe 
and $^{124}$Xe is even more intriguing: a single particle triggers the transition 
from the SD to HD minima (see Table \ref{HD-table}). Considering the fact that the 
ridges corresponding to the SD and HD minima are observed in neighboring nuclei, it 
is difficult to understand why the ridges corresponding to both minima have not 
been seen in the same nucleus.

  The calculated kinematic and dynamic moments of inertia as well as transition 
quadrupole moments of the lowest HD solutions in the candidate HD nuclei are shown 
in Fig.\ \ref{HD-exp}. The calculated $J^{(2)}$ moments of inertia somewhat 
underestimate experimental data. The results of the MM calculations for $^{118}$Te, 
$^{124}$Xe and $^{125}$Cs (see Table \ref{HD-table}) are closer to experimental 
data, but they are obtained at fixed quadrupole deformation $\beta_2$ while other 
deformation parameters $\beta_4$, $\beta_6$ and $\beta_8$ are automatically 
readjusted so as to minimize the total free Routhian for the vacuum configuration.

%%%%%%%%%%%%%%%%%%%%%%%%%%%%%%%%%%%%%%%%%%%%%%%%%%%%%%%%%%%%%%%%%%%%%%%%%%%%%%%%%
\begin{figure}[h]
\vspace{0.3cm}
\includegraphics[angle=0,width=5.0cm]{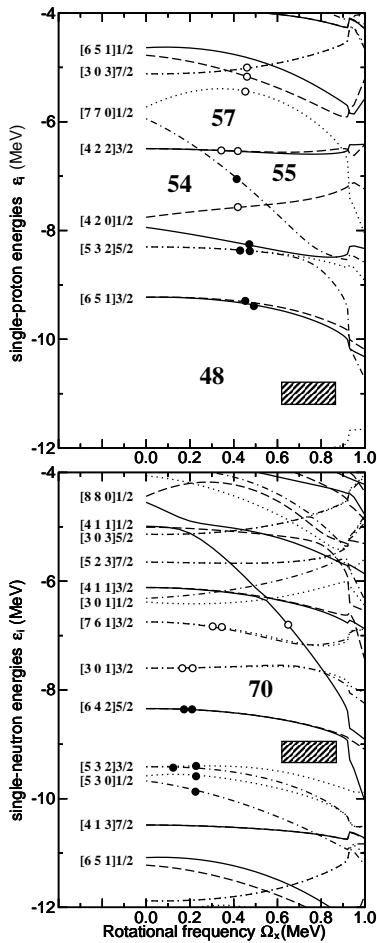}
\caption{Proton (top panel) and neutron (bottom panel) single-particle 
energies (routhians) in the self-consistent rotating potential as a 
function of the rotational frequency $\Omega_x$. They are given along 
the deformation path of the yrast HD configuration (the [1,2] conf. in Fig. \ 
\ref{Xe124-eld-NL1}) 
in $^{124}$Xe and obtained in the calculations with the NL1 parametrization 
of the RMF Lagrangian. Long-dashed, solid, dot-dashed and dotted lines indicate  
$(\pi=+, r=+i)$, $(\pi=+, r=-i)$, $(\pi=-, r=+i)$ and $(\pi=-, r=-i)$ orbitals, 
respectively. At $\Omega_x=0.0$ MeV, the single-particle orbitals are labeled by 
the asymptotic quantum numbers $[Nn_z\Lambda]\Omega$ (Nilsson quantum numbers) of 
the dominant component of the wave function. Solid (open) circles indicate the 
orbitals occupied (emptied) in the [1,2] configuration. The dashed box indicates 
the frequency range corresponding to the spin-range $I=60-85\hbar$ in this 
configuration.}
\label{Xe124-routh}
\end{figure}
%%%%%%%%%%%%%%%%%%%%%%%%%%%%%%%%%%%%%%%%%%%%%%%%%%%%%%%%%%%%%%%%%%%%%%%%%%%%%%%

 In the MM calculations, the kinematic moments of inertia of the configurations 
in the HD minimum decrease smoothly with the spin, while their dynamic moments of 
inertia are nearly constant (see Figs. 10 and 11 in Ref.\ \cite{SDH.07}). The 
behaviour of these observables as a function of rotational frequency (or spin) is 
completely different in the self-consistent CRMF calculations (see Figs. \ref{HD-exp}, 
\ref{Xe124-J2Qt} and Fig.\ \ref{near-Xe124-J2} below). The 
kinematic moment of inertia is either nearly constant or very gradually increases
with rotational frequency. The dynamic moment of inertia gradually increases over 
the calculated frequency range showing the features typical to the SD bands in the 
$A\sim 190$ mass region which are affected by pairing \cite{WS.95,BHN.95}: this is 
despite the fact that pairing is neglected in the CRMF calculations. The transition 
quadrupole moment $Q_t$ is also increasing with rotational frequency; such a feature 
has not been seen before in the calculations without pairing for the SD bands. The 
microscopic origin of these unusual features will be discussed in more details in 
Sect.\ \ref{Xe124-neigh}.

%%%%%%%%%%%%%%%%%%%%%%%%%%%%%%%%%%%%%%%%%%%%%%%%%%%%%%%%%%%%%%%%
\subsection{$^{124}$Xe nucleus}
\label{Xe124-neigh}
%%%%%%%%%%%%%%%%%%%%%%%%%%%%%%%%%%%%%%%%%%%%%%%%%%%%%%%%%%%%%%%%

   The results of the CRMF calculations for some HD 
configurations in $^{124}$Xe are displayed in Fig.\ \ref{Xe124-eld-NL1}. 
The HD minimum becomes lowest in energy at spin $82\hbar$, and the [1,2]
configuration is the yrast HD configuration in the spin range of 
interest. The occupation of the single-particle orbitals in this 
configuration is presented in Fig.\ \ref{Xe124-routh}. The excited 
HD configurations displayed in Fig.\ \ref{Xe124-eld-NL1} are built from 
this configuration by exciting either one proton or one neutron or 
simultaneously one proton and one neutron. The total number of excited HD 
configurations shown is 35. It interesting to mention that the configuration 
involving the lowest $N=8$ neutron orbital (the [1,21] conf. in Fig.\ 
\ref{Xe124-eld-NL1}) is calculated at low excitation energy.

\begin{figure}[ht]
\vspace{0.7cm}
\includegraphics[angle=0,width=9cm]{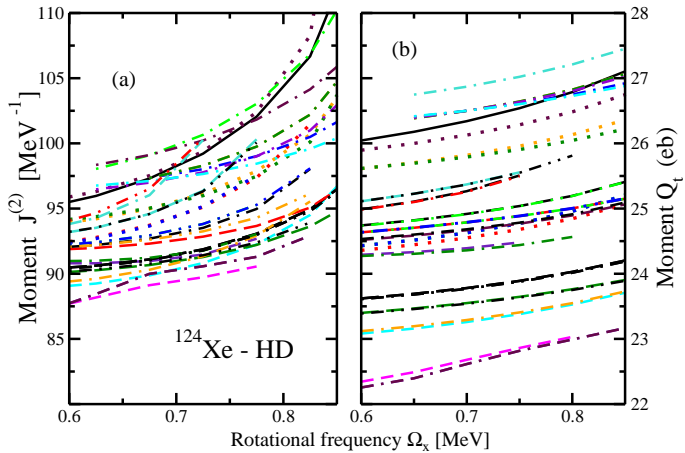}
\caption{(Color online) Dynamic moments of inertia $J^{(2)}$ (panel (a)) and transition 
quadrupole moments $Q_t$ (panel (b)) of the HD configurations in $^{124}$Xe shown 
in Fig.\ \ref{Xe124-eld-NL1}. They are displayed as a function of rotational 
frequency $\Omega_x$. The regions of band crossings are excluded in these plots.}
\label{Xe124-J2Qt}
\end{figure}

 The calculations reveal a high density of the HD configurations which will be 
even higher if the additional calculations for the excited configurations would 
be performed starting from the low-lying
excited HD configurations, such as the [1,21] configuration. This high 
density is due to two facts: relatively small $Z=54$ and $N=70$ HD shell 
gaps in the frequency range of interest (see Fig.\ \ref{Xe124-routh}) and the
softness of the potential energy surfaces in the HD minimum. Fig.\ 
\ref{Xe124-J2Qt}b illustrates the latter feature: the particle-hole excitations 
discussed above, characterised by low excitation energy, lead to appreciable changes 
in the transition quadrupole moments $Q_t$. It is interesting to mention that 
there are large similarities between the single-particle routhians in the 
vicinity of the $Z=54$ and $N=70$ HD shell gaps obtained in the CRMF calculations 
for yrast HD configuration in $^{124}$Xe (Fig.\ \ref{Xe124-routh}) and the ones 
obtained in the  Woods-Saxon calculations for the HD minimum in $^{122}$Xe employing 
the so-called universal parametrization of the Woods-Saxon potential (see Figs.\ 8 
and 9 in Ref.\ \cite{SDH.07}). As a consequence, the high density of the excited 
HD states in $^{124}$Xe is also expected in the MM calculations based on the 
formalism of Ref.\ \cite{SDH.07}.

%%%%%%%%%%%%%%%%%%%%%%%%%%%%%%%%%%%%%%%%%%%%%%%%%%%%%%%%%%%%%%%%%%%%%%%%%%%%%%%%%%
%---------------------------------------------------------------------------

  The high density of the HD configurations may question our neglect of pairing.
This is because there are numerous possibilities to scatter proton and neutron pairs 
and this process is energetically inexpensive due to the high 
density of the calculated configurations. In order to test the impact of pairing on the 
moments of inertia and binding energies, the comparative studies of the 
vacuum HD configuration and its unpaired analog in $^{124}$Xe and of the 
vacuum SD configuration and its unpaired analog in $^{152}$Dy have been 
performed within the cranked relativistic Hartree+Bogoliubov (CRHB) \cite{CRHB} and 
CRMF approaches. An approximate particle number projection by means of the Lipkin-Nogami 
method is employed in the CRHB approach. Note that unpaired analog of the vacuum HD 
configuration in $^{124}$Xe (built from the [1,2] configuration by the excitation of 
the proton from the $\pi[770]1/2(r=+i)$ orbital into the $\pi [420]1/2(r=+i)$ orbital,
see Fig.\ \ref{Xe124-routh}) is non-yrast in the spin range of interest. As follows 
from this study, in both nuclei the pairing has a similar impact on the moments of 
inertia of the configurations under consideration. Taking into account that the SD 
bands in the $A\sim 150$ mass region are well 
described in the calculations without pairing \cite{A150,ALR.98}, it is reasonable 
to expect that the neglect of pairing is a valid approximation for the moments
of inertia of the HD bands in $^{124}$Xe.  Pairing leads to an additional binding 
of $\sim 500$ keV in the case of yrast SD band in $^{152}$Dy; this additional binding 
slightly exceeds 1 MeV in the case of the vacuum HD configuration in $^{124}$Xe.  The 
dominant effects in the quenching of pairing correlations are the Coriolis antipairing 
effect and the quenching due to shell gaps: the latter effect being more pronounced 
in the SD bands of the $A\sim 150$ mass region because of the larger size of the SD 
shell gaps (see Fig.\ 4  in Ref. \cite{A150}). 
The third mechanism of the decrease of pairing is the 
blocking  effect \cite{Ring-book}. Due to this effect the impact of pairing on physical 
observables will be even lower in the HD bands of $^{124}$Xe based on the excitation(s) 
of one (two) particles considered in Fig.\ \ref{Xe124-eld-NL1}. Thus, although weak 
pairing will somewhat modify the relative energies of different configurations, in no 
way will it create an energy gap between the vacuum and excited configurations.

%%%%%%%%%%%%%%%%%%%%%%%%%%%%%%%%%%%%%%%%%%%%%%%%%%%%%%%%%%%%%%%%%%%%%%%%%%%%%%%%%%
\begin{figure}[ht]
\includegraphics[angle=0,width=8.0cm]{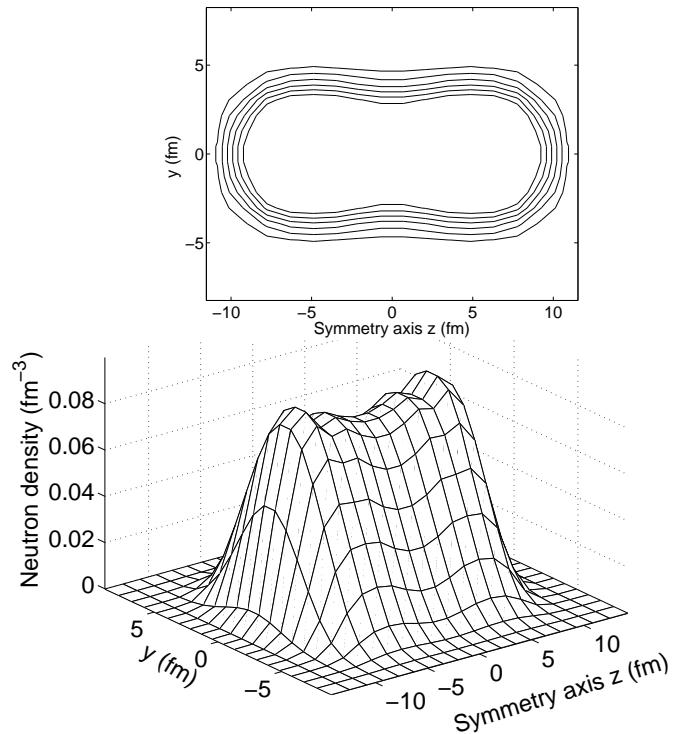}
\vspace{0.4cm}
\caption{The self-consistent neutron density $\rho_n(y,z)$ as a function 
of $y$- and $z$- coordinates for the [1,2] configuration in $^{124}$Xe at 
rotational frequency $\Omega_x=0.75$  MeV. Top and bottom panels show 2- and 
3-dimensional plots of the density distribution, respectively. In the top panel, 
the densities are shown in steps of 0.01 fm$^{-3}$ starting from 
$\rho_n(y,z)=0.01$ fm$^{-3}$.}
\label{density-Xe124}
\end{figure}
%%%%%%%%%%%%%%%%%%%%%%%%%%%%%%%%%%%%%%%%%%%%%%%%%%%%%%%%%%%%%%%%%%%%%%%%%%%%%%%%%%

%%%%%%%%%%%%%%%%%%%%%%%%%%%%%%%%%%%%%%%%%%%%%%%%%%%%%%%%%%%%%%%%%%%%%%%%%%%%%%%%%%%
\begin{figure}[ht]
\includegraphics[angle=0,width=8.0cm]{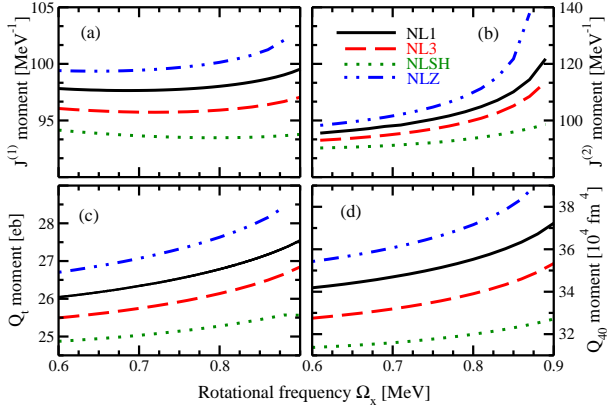}
\vspace{0.2cm}
\caption{(Color online) Kinematic ($J^{(1)}$) and dynamic ($J^{(2)}$)  moments of 
inertia as  well as transition quadrupole $Q_t$ and mass hexadecupole $Q_{40}$ moments 
of the [1,2] configuration in $^{124}$Xe calculated with different parametrizations of 
the RMF Lagrangian.}
\label{Xe124-j1-j2}
\end{figure}
%%%%%%%%%%%%%%%%%%%%%%%%%%%%%%%%%%%%%%%%%%%%%%%%%%%%%%%%%%%%%%%%%%%%%%%%%%%%%%%%%%%

 The calculations suggest that it will be difficult to observe discrete HD bands 
in $^{124}$Xe since their high density will lead to a situation in which the feeding 
intensity will be redistributed among many low-lying bands, thus drastically reducing 
the intensity with which each individual band is populated. 
On the other hand, the high density 
of the HD bands may favor the observation of the rotational patterns in the form of 
ridge-structures in three-dimensional rotational mapped spectra as it has been seen 
in the HLHD experiment \cite{Hetal.06}.

   Fig.\ \ref{HD-exp} shows that the HD shapes undergo a centrifugal stretching that 
result in an increase of the transition quadrupole moments $Q_t$ with
increasing rotational frequency. This process also reveals itself in the moments of inertia: 
the kinematic moments of inertia are either nearly constant or slightly increase with 
increasing rotational frequency, while the dynamic moments of inertia increase 
continuously and substantially over the frequency region of interest. On the contrary, 
the dynamic moments of inertia of the HD bands are almost constant as a function of 
rotational frequency in the MM calculations (see Figs. 10 and 20 in Ref.\ \cite{SDH.07}), 
which is most likely a consequence of fixed quadrupole deformation. The above mentioned 
features are general ones for the HD bands in the $A\sim 120$ mass region, see Figs.\ 
\ref{HD-exp}, \ref{Xe124-J2Qt} and \ref{near-Xe124-J2}. They are in complete contract 
to the features of the SD bands in unpaired regime, in which the $Q_t$, $J^{(1)}$ and 
$J^{(2)}$ values (apart from the unpaired band crossing regions) decrease with 
increasing rotational frequency (see Refs.\ \cite{BRA.88,A150,A60,VRAL.05}  and 
references therein).

   Systematic analysis of the yrast/near-yrast HD configurations in the part of the 
nuclear chart under investigation shows that the centrifugal stretching is a general 
feature.  At the spins, where the HD minimum is lowest in energy, it reveals 
itself (with very few exceptions) 
by the increase of transition quadrupole $Q_t$ and mass hexadecapole $Q_{40}$ moments.
Only in a few HD bands, characterized by the modest transition quadrupole moment, 
at low rotational frequencies these quantities decrease with increasing $\Omega_x$. 
However, even in these bands the $Q_t$ and $Q_{40}$ values start to increase above 
specific value of rotational frequency. Similar features are also seen in the dynamic 
moments of inertia; with a few exceptions the $J^{(2)}$ values increase in the spin 
range of interest. The variations (both the increases and decreases) in the kinematic
moments of inertia are rather small ($\sim 2\%$ of absolute value) in the frequency 
range of interest.

%%%%%%%%%%%%%%%%%%%%%%%%%%%%%%%%%%%%%%%%%%%%%%%%%%%%%%%%%%%%%%%%%%%%%%%%%%%%%%%%%%
\begin{figure}[h]
\vspace{1.0cm}
\includegraphics[angle=0,width=8.0cm]{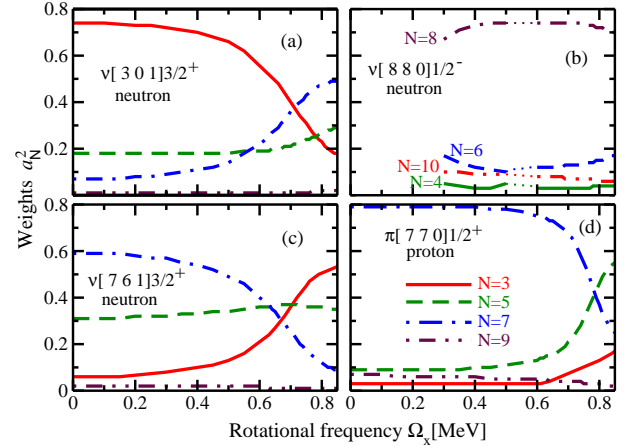}
\vspace{0.4cm}
\caption{The weights $a_N^2$ of different $N$-components in the structure of the 
wave functions of the indicated orbitals. They are shown as a function of rotational 
frequency. For simplicity, the region of the crossing between the $\nu[880]1/2^-$ 
and $\nu[411]3/2^-$ orbitals at $\Omega_x \sim 0.55$ MeV is removed; dotted lines are 
used in panel (b) to connect the weights corresponding to the $\nu[880]1/2^-$ 
orbital before and after crossing.}
\label{wave-function-Xe124}
\end{figure}
%%%%%%%%%%%%%%%%%%%%%%%%%%%%%%%%%%%%%%%%%%%%%%%%%%%%%%%%%%%%%%%%%%%%%%%%%%%%%%%%%%

%%%%%%%%%%%%%%%%%%%%%%%%%%%%%%%%%%%%%%%%%%%%%%%%%%%%%%%%%%%%%%%%%%%%%%%%%%%%%%%%%%
\begin{figure*}[ht]
\includegraphics[angle=0,width=13.0cm]{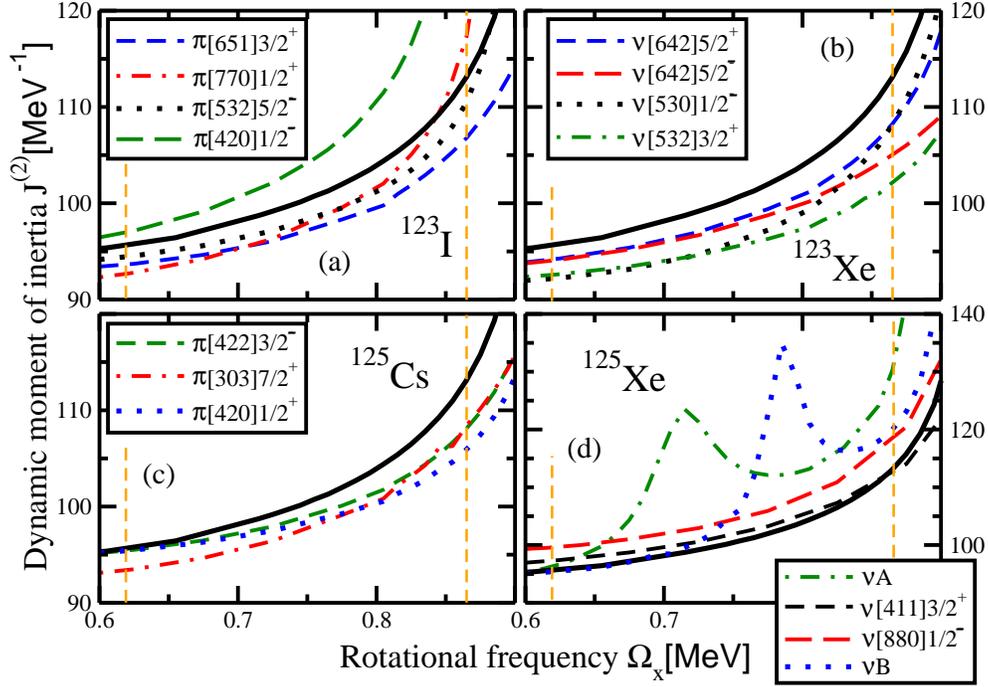}
\vspace{0.4cm}
\caption{(Color online) Dynamic moments of inertia $J^{(2)}$ of selected configurations
in $^{124}$Xe and neighbouring nuclei.  Dynamic moment of inertia of the 
[1,2] configuration A in $^{124}$Xe is shown by a thick solid line in each 
panel. The $J^{(2)}$ values of the configurations in the nucleus indicated
on the panel are displayed by the lines of other types. These 
configurations differ from the [1,2] configuration A in $^{124}$Xe in the 
occupation of the orbitals shown in the panels. Vertical dashed lines indicate 
the frequency range corresponding to the spin range $I=60 - 85\hbar$ in the 
[1,2] configuration of $^{124}$Xe. } 
\label{near-Xe124-J2}
\end{figure*}
%%%%%%%%%%%%%%%%%%%%%%%%%%%%%%%%%%%%%%%%%%%%%%%%%%%%%%%%%%%%%%%%%%%%%%%%%%%%%%%%%%

 The basis of the CRMF model is sufficiently large to see if there is a tendency 
for the development of necking. Fig.\ \ref{density-Xe124} shows some indications 
of the necking and the clusterization of the density into two fragments in the 
[1,2] configuration of $^{124}$Xe, but this effect is not very pronounced in this
nucleus.

    The kinematic and dynamic moments of inertia as well as the transition quadrupole
and mass hexadecapole moments of the [1,2] configuration in $^{124}$Xe are shown 
for different parametrizations of the RMF Lagrangian in Fig.\ \ref{Xe124-j1-j2}. 
The gradual increase of all physical observables is due to centrifugal stretching.
The NLZ (NLSH) parametrizations provide the largest (smallest) values of the above 
mentioned physical observables, while  the results obtained with NL1 and NL3 are 
in between those results. Similar relations between the results obtained with 
these parametrizations also exist in other regions of nuclear chart studied so 
far in the CRMF or CRHB frameworks, namely, in the $A\sim 60$ \cite{ARR.99}, 
$A\sim 150$ \cite{ALR.98} and $A\sim 190$ \cite{CRHB} regions of superdeformation 
and in the $A\sim 250$ \cite{A250} region of normal deformation. The NL1 and 
NL3 parametrizations, which have been extensively used in the previous studies of 
rotating systems and superdeformation \cite{VRAL.05}, give the values of physical 
observables of interest which differ only by few \%. It is known that the NLSH 
parametrization somewhat underestimates the experimental moments of inertia 
\cite{ARR.99,ALR.98}. The NLZ parametrization has not been used in the previous 
studies of rotating systems, so it is unknown how well it describes such systems.

%%%%%%%%%%%%%%%%%%%%%%%%%%%%%%%%%%%%%%%%%%%%%%%%%%%%%%%%%%%%%%%%%%%%%%%
\subsection{Single-particle properties at hyperdeformation: an example
of neighbourhood of $^{124}$Xe.}
\label{Xe124-single}
%%%%%%%%%%%%%%%%%%%%%%%%%%%%%%%%%%%%%%%%%%%%%%%%%%%%%%%%%%%%%%%%%%%%%%%%

  The role of the single-particle degrees of freedom at hyperdeformation 
was mainly overlooked in the previous studies. It has been studied to 
some extent only within the MM method in Refs.\ \cite{A.93,SDH.07}. However, 
the studies of Ref.\ \cite{SDH.07} suggest that the $^{124}$Xe nucleus is 
very rigid in the HD minimum: the dynamic moments of inertia of different HD 
bands differ by no more than 2\%, and their changes as a function of spin 
are very small (see Fig.\ 10 in Ref.\ \cite{SDH.07}). Similar results 
were obtained for HD bands in $^{146}$Gd and $^{152}$Dy in Ref.\ \cite{A.93}.

%%%%%%%%%%%%%%%%%%%%%%%%%%%%%%%%%%%%%%%%%%%%%%%%%%%%%%%%%%%%%%%%%%%%%%%%%%%%%%%%%%%
\begin{figure*}[ht]
\includegraphics[angle=0,width=13.0cm]{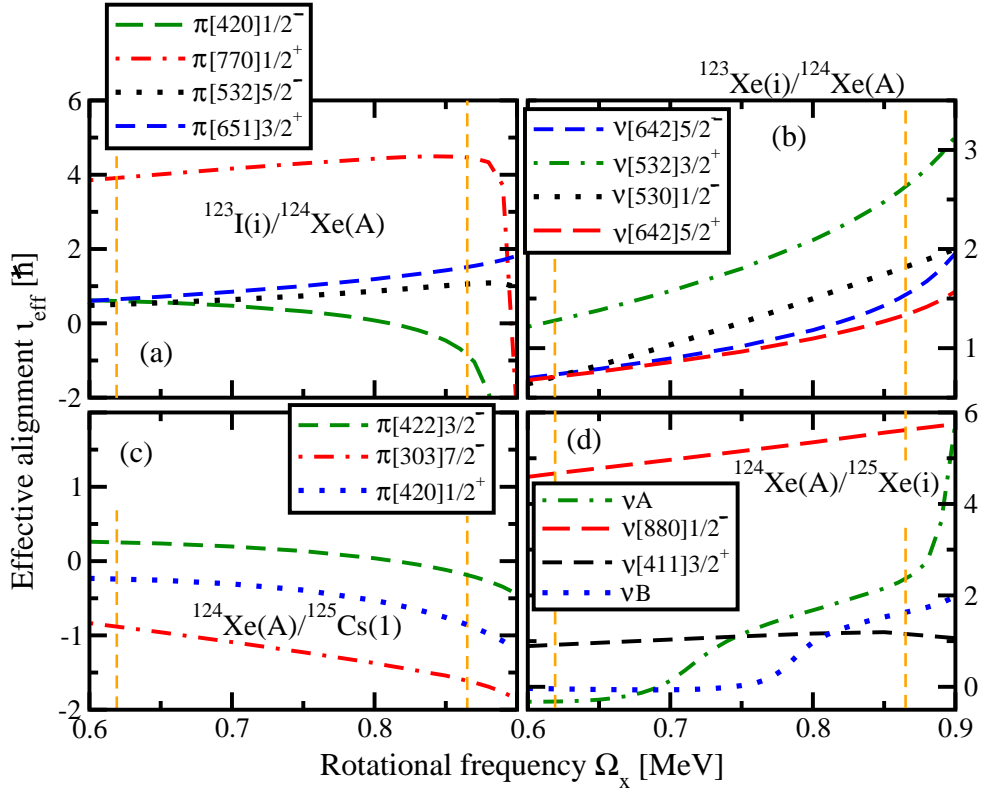}
\vspace{0.2cm}
\caption{(Color online) Effective alignments $i_{eff}$ extracted from the calculated 
configurations for the orbitals active in the vicinity of the $Z=54/55$ and $N=70$ 
HD shell  gaps (see Fig.\ \ref{Xe124-routh}). The calculated configurations are 
the [1,2] conf. in $^{124}$Xe and the configurations in neighboring nuclei (shown 
in Fig.\ \ref{near-Xe124-J2}) obtained by adding or removing a single particle 
(proton or neutron). The effective alignment between configurations X and Y is 
indicated as ``X/Y''. The configuration X in the lighter nucleus is taken as a 
reference, so the effective alignment measures the effect of the additional particle. 
The compared configurations differ in the occupation of the orbitals shown in the 
panels. Note that the vertical scale of different panels is different. Vertical 
dashed lines indicate the frequency range corresponding to the spin range 
$I=60 - 85\hbar$ in the conf. A of $^{124}$Xe.}
\label{Xe124-ief}
\end{figure*}
%%%%%%%%%%%%%%%%%%%%%%%%%%%%%%%%%%%%%%%%%%%%%%%%%%%%%%%%%%%%%%%%%%%%%%%%%%%%%%%%%%%

   On the contrary, the CRMF calculations for the dynamic moment of inertia
of the yrast and excited HD configurations in $^{124}$Xe show much larger 
spread and much larger variations as a function of rotational frequency, see 
Fig.\ \ref{Xe124-J2Qt}a. In addition, large variations in the calculated 
transition quadrupole moments $Q_t$ of these configurations are clearly seen
in Fig.\ \ref{Xe124-J2Qt}b. This suggests that the HD minimum is relatively
soft and that the individual properties of the single-particle orbitals play
an important role in the definition of the properties of the HD bands. 
One of our goals is to investigate the impact of the particle in a specific 
single-particle orbital on the  properties of the HD bands and to study 
whether the methods of configuration assignment based on the relative 
properties of different bands are also applicable at HD.

%%%%%%%%%%%%%%%%%%%%%%%%%%%%%%%%%%%%%%%%%%%%%%%%
\subsubsection{The structure of the wave function}
%%%%%%%%%%%%%%%%%%%%%%%%%%%%%%%%%%%%%%%%%%%%%%%%

   The structure of the  wave function at HD is analysed on the example of a few 
single-particle orbitals of the [1,2] configuration in $^{124}$Xe (Fig.\ \ref{wave-function-Xe124}). 
The evolution of these orbitals in energy with rotational frequency is displayed in Fig.\ 
\ref{Xe124-routh}. The wave function $\Psi$ is expanded into the basis states by
\begin{eqnarray}
\Psi = \sum_{N,\alpha} c_{N,\alpha} |N \alpha> 
\end{eqnarray}
where $N$ and $\alpha$ represent the principal quantum number and the set of
additional quantum numbers specifying the basis state, respectively.  We specify the 
weight $a_N^2$ of the basis states belonging to the specific value of $N$ 
in the structure of the wave function as 
\begin{eqnarray}
a_N^2 = \sum_{N-{\rm fixed}, \alpha} c^2_{N,\alpha}
\end{eqnarray}
with the condition  $\sum_{N} a_N^2 =1$ following from the orthonormalization
of the wave function of the single-particle orbital. 
  
  Hyperdeformation leads to a considerable fragmentation of 
the wave function over $N$, which is much larger than in
the case of SD. In the regions away from the band crossing the weight 
$a_N^2$ of the dominant $N$-component of the wave function does not exceed 0.8 while
the weight of second largest component is typically around 0.2 (Fig.\ 
\ref{wave-function-Xe124}). Very strong fragmentation of the wave function is seen in 
the case of the $\nu [761]3/2^+$ orbital: before the band crossing the weights of 
the $N=7$ and $N=5$ components of the wave function are approximately 0.6 and 0.3, 
respectively. Even stronger fragmentation is seen in the region of the band 
crossing of the $\nu [761]3/2^+$ and  $\nu [301]3/2^+$ orbitals  at $\Omega_x \sim 0.7$ 
MeV (Figs.\ \ref{Xe124-routh}) where they strongly interact and gradually exchange 
their character (Figs.\ \ref{wave-function-Xe124}a and c). Similar fragmentation 
is also seen for the $\pi [770]1/2^+$ orbital (Fig.\ \ref{wave-function-Xe124}) 
which interacts strongly with the $\pi [532]5/2^+$ orbital in the band crossing 
region at $\Omega_x \sim 0.8$ MeV (Fig.\ \ref{Xe124-routh}).

%%%%%%%%%%%%%%%%%%%%%%%%%%%%%%%%%%%%%%%%%%%%%%%%%%%%%%%%%%%%%%%%%%%%%%%%%%%%%%%%%%
\begin{figure*}[ht]
\includegraphics[angle=0,width=13.0cm]{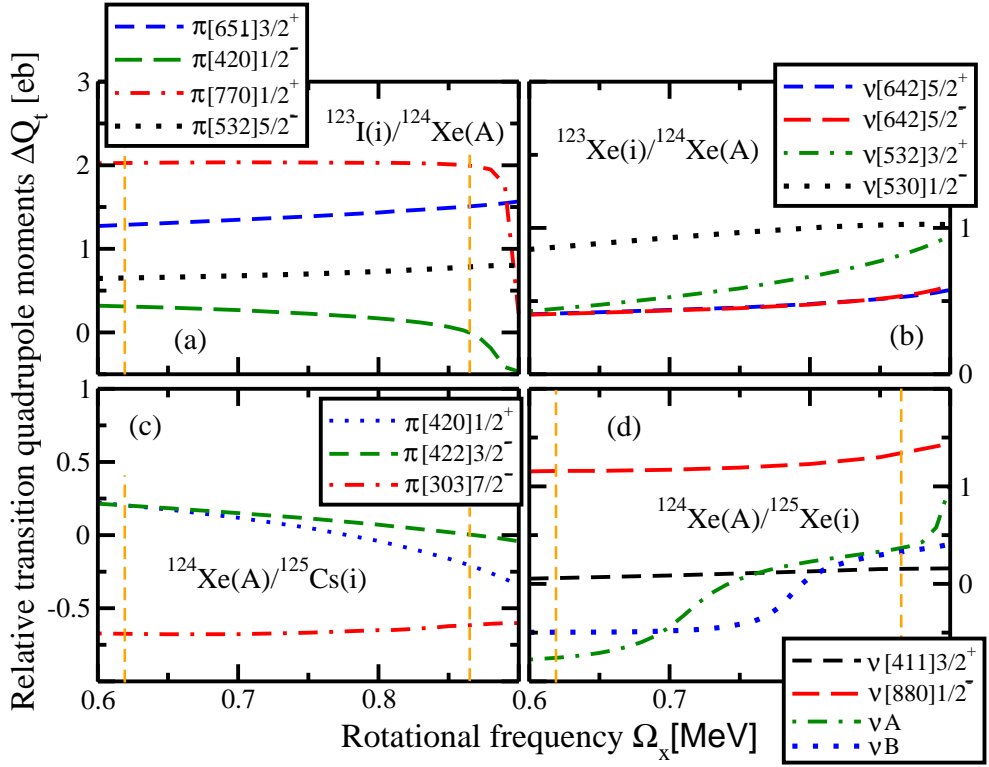}
\vspace{0.4cm}
\caption{ (Color online) Relative transition quadrupole moments 
$\Delta Q_t = Q_t (A+1) - Q_t (A)$ [$A$ is the mass of the nucleus] extracted 
from the calculated configurations in indicated nuclei. The compared configurations 
are shown as ``X/Y'': the configuration X in the lighter nucleus is taken as a 
reference, so the $\Delta Q_t$ measures the effect of the additional particle 
placed in the  orbitals shown in the panels. Vertical dashed lines indicate the 
frequency range corresponding to the spin range $I=60 - 85\hbar$ in the [1,2] 
configuration of $^{124}$Xe. }
\label{near-Xe124-Qt}
\end{figure*}
%%%%%%%%%%%%%%%%%%%%%%%%%%%%%%%%%%%%%%%%%%%%%%%%%%%%%%%%%%%%%%%%%%%%%%%%%%%%%%%%%%

%  At $\Omega_x = 0.0$ MeV the weight $c^2_{N\alpha}$ of the dominant component of the wave 
% function is larger than 0.5 for the majority of the orbitals. This fact allows to use 
% the asymptotic quantum numbers $[Nn_z\Lambda]\Omega$ (Nilsson quantum numbers 
% \cite{NR.book}) for the labeling of the single-particle orbitals. It is necessary, 
% however, to remember that with increasing rotational frequency, the fragmentation 
% of the wave function increases with a corresponding decrease (frequently drastic) 
% of the weight $c^2_{N\alpha}$  of the dominant component of the wave function.

%%%%%%%%%%%%%%%%%%%%%%%%%%%%%%%%%%%%%%%%%%%%%%%%%%%%%%%%%
\subsubsection{The methods of configuration assignment}
%%%%%%%%%%%%%%%%%%%%%%%%%%%%%%%%%%%%%%%%%%%%%%%%%%%%%%%%%

  The HD bands in nuclei neighboring to $^{124}$Xe, which differ by either one 
proton or one neutron from the [1,2] configuration in $^{124}$Xe, and their 
relative properties with respect of the [1,2] configuration in  $^{124}$Xe 
are studied in order to investigate the applicability of different methods 
of configuration assignment at HD.

  The dynamic 
moments of inertia for the four HD bands in each of these nuclei are compared with 
the one of the [1,2] configuration in $^{124}$Xe in Fig.\ \ref{near-Xe124-J2}. The 
difference between the dynamic moments of inertia of the configurations in nuclei 
with masses $A$ and $A\pm 1$ is due to the impact of the particle in the specific 
single-particle orbital by which two compared configurations differ. 
The results of the calculations question conventional wisdom \cite{BRA.88} that 
the largest impact on the dynamic moment of inertia is coming from the particles in 
the intruder orbitals. Indeed, the impact of the neutron in the hyperintruder $\nu [880]1/2^-$ 
orbital on the dynamic moments of inertia (Fig.\ \ref{near-Xe124-J2}d) is comparable
to the one of non-intruder $\nu [642]5/2^+$ orbital or even smaller by a factor of 
$\sim 2$ than the impact due to the neutron in non-intruder $\nu [532]3/2^+$ orbital (Fig.\ 
\ref{near-Xe124-J2}b). A similar situation is also seen for protons, where, for example, 
the impact  of the proton in the hyperintruder $\pi [770]1/2^+$ orbital is smaller than 
its impact in the non-intruder $\pi [420]1/2^-$ orbital. This suggests that not only 
angular momentum, carried by the particle in specific single-particle orbital, but 
also polarization effects it induces into time-even and time-odd mean fields 
\cite{AR.00} are important when considering relative properties of two configurations. 
Based on this example, one can conclude that the configuration assignment of the HD bands, 
based only on the relative properties of the dynamic moments of inertia of two compared 
bands, is unreliable.

%%%%%%%%%%%%%%%%%%%%%%%%%%%%%%%%%%%%%%%%%%%%%%%%%%%%%%%%%%%%%%%%%%%%%%%%%%%%%%%%%%%
\begin{figure*}[ht]
\includegraphics[angle=0,width=16.0cm]{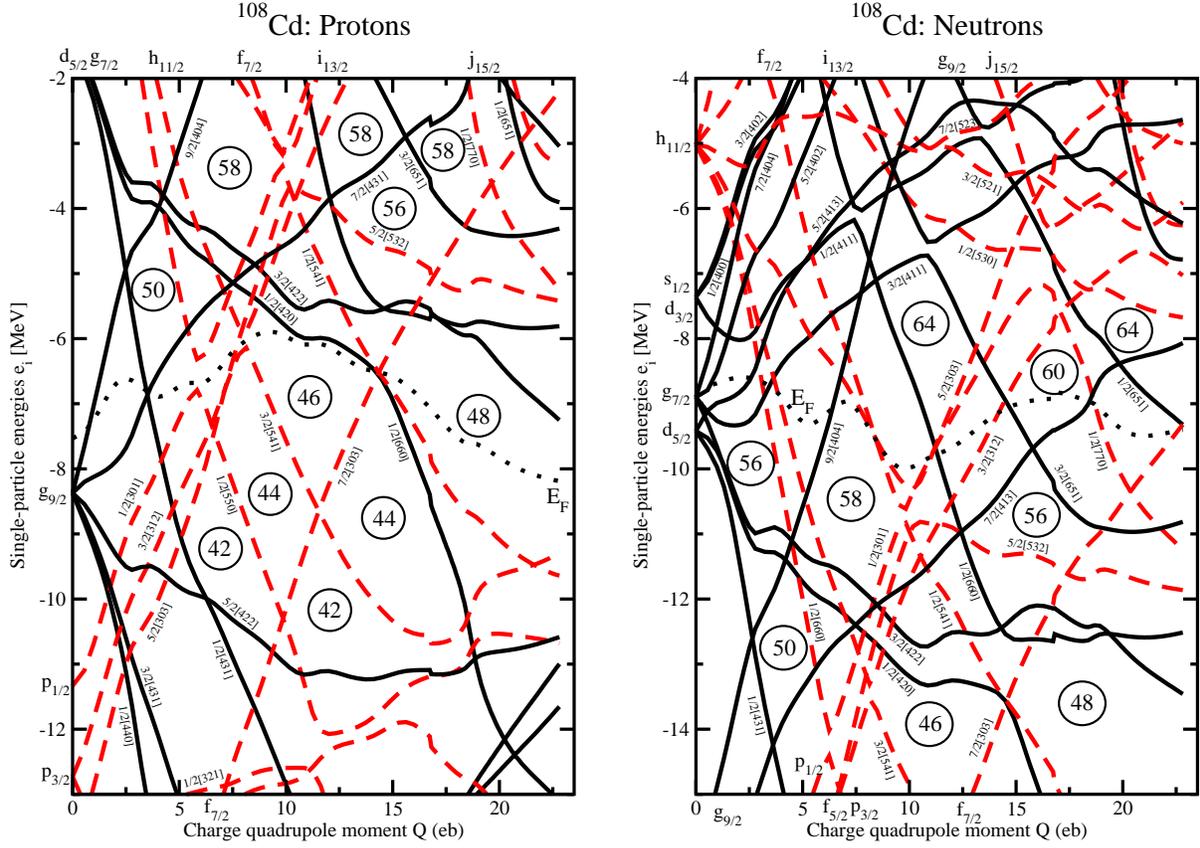}
\vspace{0.2cm}
\caption{(Color online) Proton and neutron single-particle energies in $^{108}$Cd, as 
a function of charge quadrupole moment $Q$, obtained in the axially symmetric RMF 
calculations. Solid and dashed lines denote positive and negative parity orbitals, 
respectively. The Fermi energy ${\rm E_F}$ is shown by dotted line. The single-particle 
orbitals are labeled by the Nilsson quantum numbers. Large shell gaps are indicated.}
\label{108Cd-nilsson}
\vspace{0.2cm}
\end{figure*}
%%%%%%%%%%%%%%%%%%%%%%%%%%%%%%%%%%%%%%%%%%%%%%%%%%%%%%%%%%%%%%%%%%%%%%%%%%%%%%%%%%%

   The configuration assignments at SD have been mostly based on the effective alignment 
approach (see Refs.\ \cite{Rag.93,ALR.98,ARR.99} and 
references therein). The success of this method is due to the fact that it was 
possible to separate intruder and non-intruder orbitals since the former show 
pronounced dependence of the effective alignments $i_{eff}$ on the rotational 
frequency (see, for example, Figs. 2, 3, 5, 6 and 8 in Ref.\ \cite{ALR.98}). On 
the contrary, the effective alignments of non-intruder orbitals are typically 
constant as a function of rotational frequency. It also follows from the
studies in the $A\sim 140-150$ region of superdeformation that the change of 
effective alignment by $\approx 1\hbar$ within the observed frequency range 
allows to identify aligning intruder orbitals with a high level of confidence.

   A  configuration assignment based on the effective alignments depends on how 
accurately these alignments can be predicted. For example, the application of the 
effective alignment approach in the $A\sim 140-150$ region of superdeformation 
requires an accuracy in the prediction of $i_{eff}$ on the level of $\sim 0.3\hbar$ 
and  $\sim 0.5\hbar$ for nonintruder and intruder orbitals, respectively 
\cite{ALR.98,Rag.93,BHN.95}. In the highly deformed and SD bands from the 
$A\sim 60-80$ mass region, these requirements for accuracy are somewhat relaxed 
\cite{ARR.99,AF.05}.  We expect that in the $A\sim 125$ mass region of HD, 
the effective alignments should be predicted with a precision similar to 
that in the $A\sim 140-150$ region for a reliable configuration assignment.

%%%%%%%%%%%%%%%%%%%%%%%%%%%%%%%%%%%%%%%%%%%%%%%%%%%%%%%%%%%%%%%%%%%%%%%%%%%%%%%%%%
\begin{figure}[ht]
\includegraphics[angle=0,width=8.0cm]{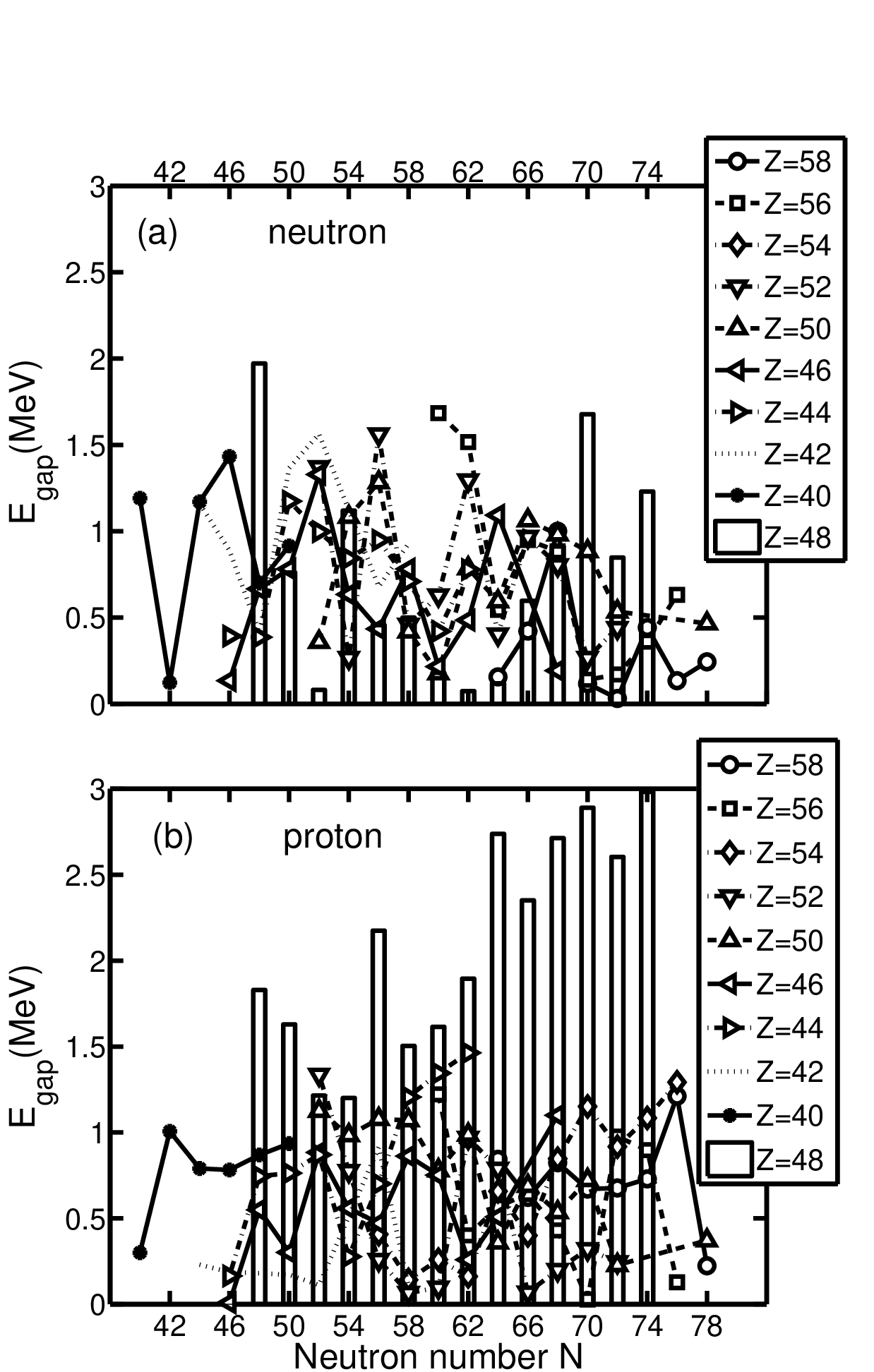}
\vspace{0.4cm}
\caption{The energy gaps between the last occupied and first unoccupied single-particle 
orbitals shown as a function of neutron number for different isotope chains.
They are extracted from the routhian diagrams of  the lowest HD configurations at 
the spin values where these configurations become yrast. The bars are used for the 
energy gaps in the Cd isotopes.}
\label{shell-gaps}
\end{figure}
%%%%%%%%%%%%%%%%%%%%%%%%%%%%%%%%%%%%%%%%%%%%%%%%%%%%%%%%%%%%%%%%%%%%%%%%%%%%%%%%%%

  Our analysis shows that a reliable configuration assignment for the HD 
bands based solely on the effective alignment approach will be problematic 
(at least in the $A\sim 125$ mass region) because of several reasons. First, 
the hyperintruder orbitals do not show appreciable variations of $i_{eff}$ with 
rotational frequency. Fig.\ \ref{Xe124-ief} shows that the effective alignments 
of the hyperintruder orbitals such as  $\pi [770]1/2^+$ and $\nu [880]1/2^-$ show 
little variations with rotational frequency (see Fig.\ \ref{Xe124-ief}a,d). On 
the contrary, the effective alignments of the $\nu [532]3/2^+$ and 
$\nu [530]1/2^-$ orbitals show much larger variations reaching $1.5\hbar$ in 
the spin range $I=60-85\hbar$ in the case of the latter orbital (see Fig.\ 
\ref{Xe124-ief}b). However, the variations of $i_{eff}$ as a function of 
rotational frequency are small for the majority of the orbitals in the spin 
range of interest.  Thus, contrary to the case of SD, it will be more difficult to 
distinguish between  hyperintruder, intruder and non-intruder orbitals based on 
the variations of $i_{eff}$ with rotational frequency.  This situation will become 
even more complicated if the suggestion of Ref.\ \cite{SDH.07} that the spin range 
over which the HD bands are expected to be observed ($24\hbar$ at the most; this is 
shorter than in the case of SD) is true. These two features (small variations of 
$i_{eff}$ and expected spin (frequency) range of the HD bands) will lead to a situation 
where the $i_{eff}$ values for many orbitals will  look alike within the typical 
'error bars' of the description of $i_{eff}$ by theoretical models,  so that it will 
be difficult to distinguish between them within  the framework of the effective 
alignment approach.

  Similar to the case of SD \cite{SDDN.96,MADLN.07}, additional 
information on how the single particle affects the properties of the HD 
bands can be extracted from the relative transition quadrupole moments 
$\Delta Q_t$. Fig.\ \ref{near-Xe124-Qt} shows that the hyperintruder 
$\pi [770]1/2^+$ and $\nu [880]1/2^-$ orbitals with $\Delta Q_t \approx 2$ $e$b 
and $\Delta Q_t \approx 1.25$ $e$b have the largest impact  on the 
transition quadrupole moments among the studied proton and neutron orbitals. 
One has to keep in mind that the addition of a proton changes the proton number 
by one. This change contributes approximately 0.5 $e$b in relative transition 
quadrupole moment $\Delta Q_t$ of the proton orbitals. This effect is not 
present in the $\Delta Q_t$ values of the neutron orbitals.

   The $\Delta Q_t$  values were used only as a complimentary tool of the 
configuration assignment at SD. This is because of the difficulty to measure 
them in experiment \cite{146Gd-exp,A130-exp1} and the fact that they show 
little variation as a function of rotational frequency, thus providing
less information than $i_{eff}$. The same features are also valid at HD; see 
Fig.\ \ref{near-Xe124-Qt} for the variations of the $\Delta Q_t$ values. In 
addition, some single-particle orbitals such as $\pi [422]3/2^-$ and 
$\pi [303]7/2^-$ (Fig.\ \ref{near-Xe124-Qt}c) show very similar $\Delta Q_t$ 
values. This will not allow to make a unique configuration assignment even 
if the experimental $\Delta Q_t$ values for these orbitals are available. On 
the other hand, their $i_{eff}$ values differ by $\sim 1\hbar$ (Fig.\ \ref{Xe124-ief}c), 
and this fact can be used in the configuration assignment.

  However, the fact that in general the effective alignment approach fails to 
provide a unique configuration assignment at HD increases the role of the 
method of configuration assignment based on relative transition quadrupole 
moments. Our analysis shows that {\it only simultaneous application of these two 
methods by comparing experimental and theoretical $(i_{eff},\Delta Q_t)$ 
values will lead to a reliable configuration assignment at HD}.

 Let us illustrate this on the hypothetical example of two ``experimental'' bands; 
one in $^{123}$I and another in $^{124}$Xe. In this example, the [1,2] configuration 
is assigned to the band in $^{124}$Xe. Let us assume that the effective alignments 
in the $^{123}$I/$^{124}$Xe pair of the bands increase from $4.0\hbar$ to $4.25\hbar$ 
in the frequency range 0.62-0.87 MeV under selected spins of these bands. 
Under these conditions, the ``experimental'' bands differ in the occupation of 
the $\pi [770]1/2^{+}$ orbital (Fig.\ \ref{Xe124-ief}a).  However, it is reasonable 
to expect that the spins of ``experimental'' bands will not be fixed, so these 
changes in effective alignment should be from $(4.0+n)\hbar$ to $(4.25+n)\hbar$, 
where $n=0,\pm 1, \pm 2, ...$.  Assuming
that the accuracy of the description of effective alignments in theoretical
calculations is around $0.4\hbar$, one can conclude that for $n=-3$ the ``experimental'' 
bands can also differ in the occupation of either the $\pi [532]5/2^-$ or $\pi [651]3/2^+$ 
orbitals (Fig.\ \ref{Xe124-ief}a). In a similar way to the $A\sim 150$ region of 
SD \cite{Rag.93,ALR.98}, the systematic studies of the pairs of the bands which differ 
by one proton may narrow the choice of the orbitals involved. On the other hand, the 
$\Delta Q_t$ values for these orbitals are drastically different; $\Delta Q_t\approx 2.0$ 
$e$b for the $\pi [770]1/2^{+}$ orbital, $\Delta Q_t \approx 1.4$ $e$b for 
$\pi [651]3/2^{+}$, and $\Delta Q_t\approx 0.7$ 
$e$b for $\pi [532]5/2^{-}$ (see Fig.\ \ref{near-Xe124-Qt}). So, if both quantities, 
$i_{eff}$ and  $\Delta Q_t$, are measured simultaneously, a unique configuration 
assignment for ``experimental'' band in $^{123}$I will be possible.

  The band crossing features of the HD bands provide an additional tool 
of configuration assignment which  can be used more frequently 
than in the case of the SD bands because of strong mixing between the 
different $N$-shells at HD. The large peaks in $J^{(2)}$ of the $\nu A$  and 
$\nu B$ configurations in $^{125}$Xe
(Fig.\ \ref{near-Xe124-J2}d) are due to the band crossings with 
a strong interaction. These crossings are also visible in the effective 
alignments $i_{eff}$ (Fig.\ \ref{Xe124-ief}d) 
and relative transition quadrupole moments $\Delta Q_t$ (Fig.\ \ref{near-Xe124-Qt}d). 
They originate from the crossing of the same signatures of the $\nu [301]3/2$ 
and $\nu [761]3/2$ orbitals, where $\nu A$ and $\nu B$ have signatures $r=+i$ 
and $r=-i$, respectively.  The former orbital is occupied before band crossing, 
the latter after band crossing. An unusual feature of these band 
crossings is the fact that they originate from the 
interaction of the orbitals, the dominant $N$-components of which differ by 
$\Delta N=4$.  At SD, the crossings between 
the orbitals dominated by different $N$-shells have been characterized by a 
weak interaction leading to a sharp jump in $J^{(2)}$ \cite{A150,Sm142b,Paris98}. 
The observed unpaired SD band crossings with strong interaction are between 
the orbitals with the same dominant $N$-shells and they were observed in the
nuclei around $^{147}$Gd \cite{R.91,A150}.

%%%%%%%%%%%%%%%%%%%%%%%%%%%%%%%%%%%%%%%%%%%%%%%%%%%%%%%%%%%%%%%%%%%%%%%
\subsection{General observations: the density of the HD bands and
the necking degree of freedom}
\label{gen-obs}
%%%%%%%%%%%%%%%%%%%%%%%%%%%%%%%%%%%%%%%%%%%%%%%%%%%%%%%%%%%%%%%%%%%%%%%

  As discussed in Sect.\ \ref{Xe124-neigh} on the example of $^{124}$Xe,
the high density of the HD bands is one of the major obstacles for 
the observation of discrete HD bands. It will lead to a situation 
where the feeding intensity will be redistributed among many low-lying HD bands, 
thus, drastically reducing the intensity with which each individual band is populated. 
As a consequence, the feeding intensity of an individual HD band will drop below the 
observational limit of experimental facility; this fact has to be taken into account 
when planning future experiments for a search of discrete HD bands.

%%%%%%%%%%%%%%%%%%%%%%%%%%%%%%%%%%%%%%%%%%%%%%%%%%%%%%%%%%%%%%%%%%%%%%%%%%%%%%%%%%
\begin{figure*}[ht]
\includegraphics[angle=0,width=17.0cm]{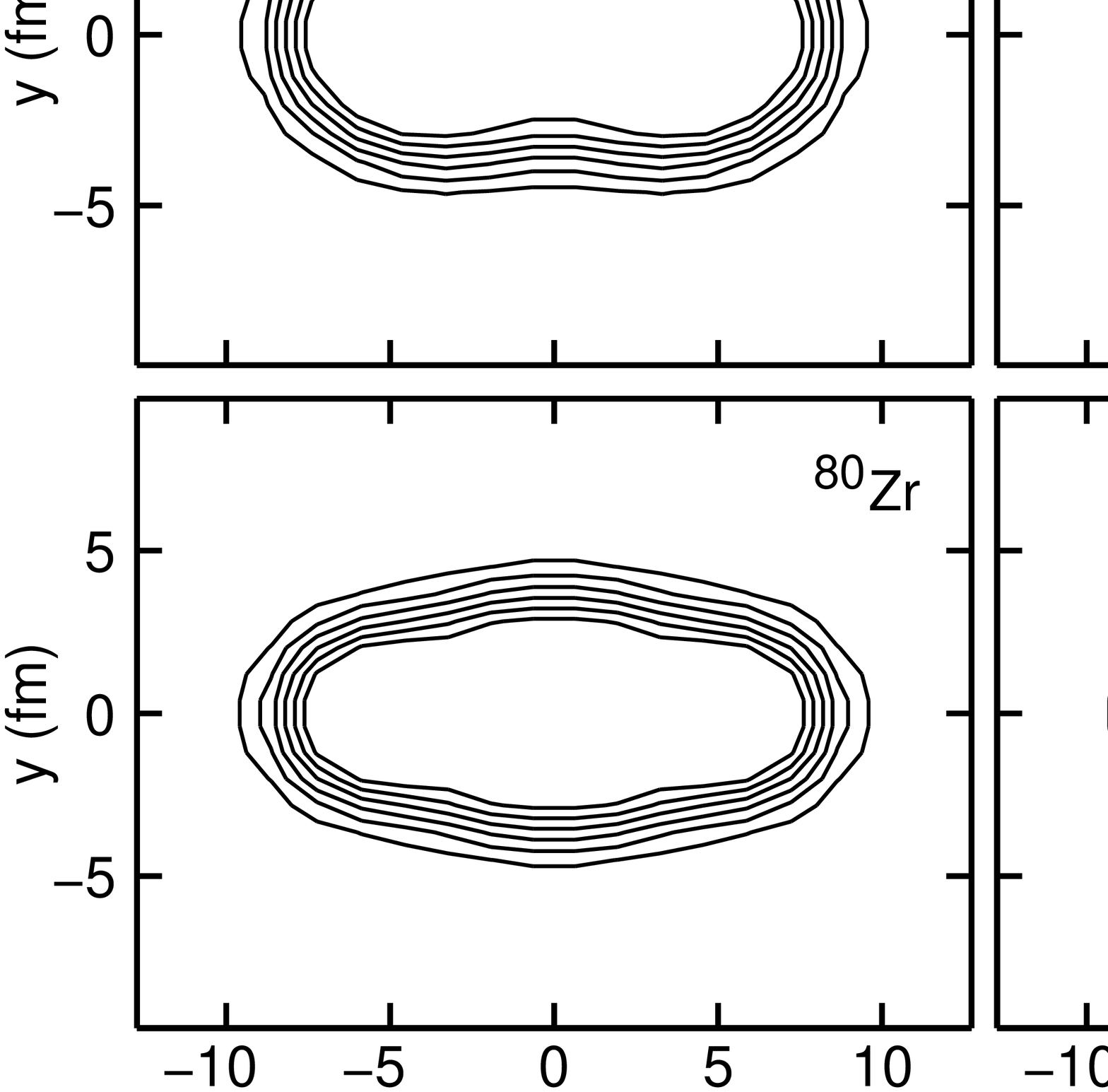}
\vspace{0.4cm}
\caption{The self-consistent proton density $\rho_p(y,z)$ as a function of 
$y$- and $z$- coordinates for the HD configurations. They are displayed at spin 
values at which these configurations become yrast. For each isotope chain, 
the densities in two nuclei (typically, most proton- and neutron rich ones included
in calculations) are shown. The densities are displayed in steps of 0.01 fm$^{-3}$ 
starting from  $\rho_p(y,z)=0.01$ fm$^{-3}$.}
\label{density-sys}
\end{figure*}
%%%%%%%%%%%%%%%%%%%%%%%%%%%%%%%%%%%%%%%%%%%%%%%%%%%%%%%%%%%%%%%%%%%%%%%%%%%%%%%%%%

  Two factors contribute to the high density of the HD bands, namely, relatively small 
proton and neutron HD shell gaps in the frequency range of interest and the softness 
of the potential energy surfaces in the HD minimum (see Sect.\ \ref{Xe124-neigh}). Systematic 
mapping of the density of the HD states as a function of the proton and neutron numbers
is too costly in the computational sense because it involves the calculation of the lowest in 
energy particle-hole excitations. Thus, we decided to look at the problem of the density of 
the HD states in a somewhat simplistic way by considering the proton and neutron energy gaps 
between the last occupied and the first unoccupied states in the yrast HD configurations; 
the small size of these gaps will most likely point to the high density of the HD bands.

%%%%%%%%%%%%%%%%%%%%%%%%%%%%%%%%%%%%%%%%%%%%%%%%%%%%%%%%%%%%%%%%%%%%%%%%%%%%%%%%%%
\begin{figure}[ht]
\includegraphics[angle=0,width=8.0cm]{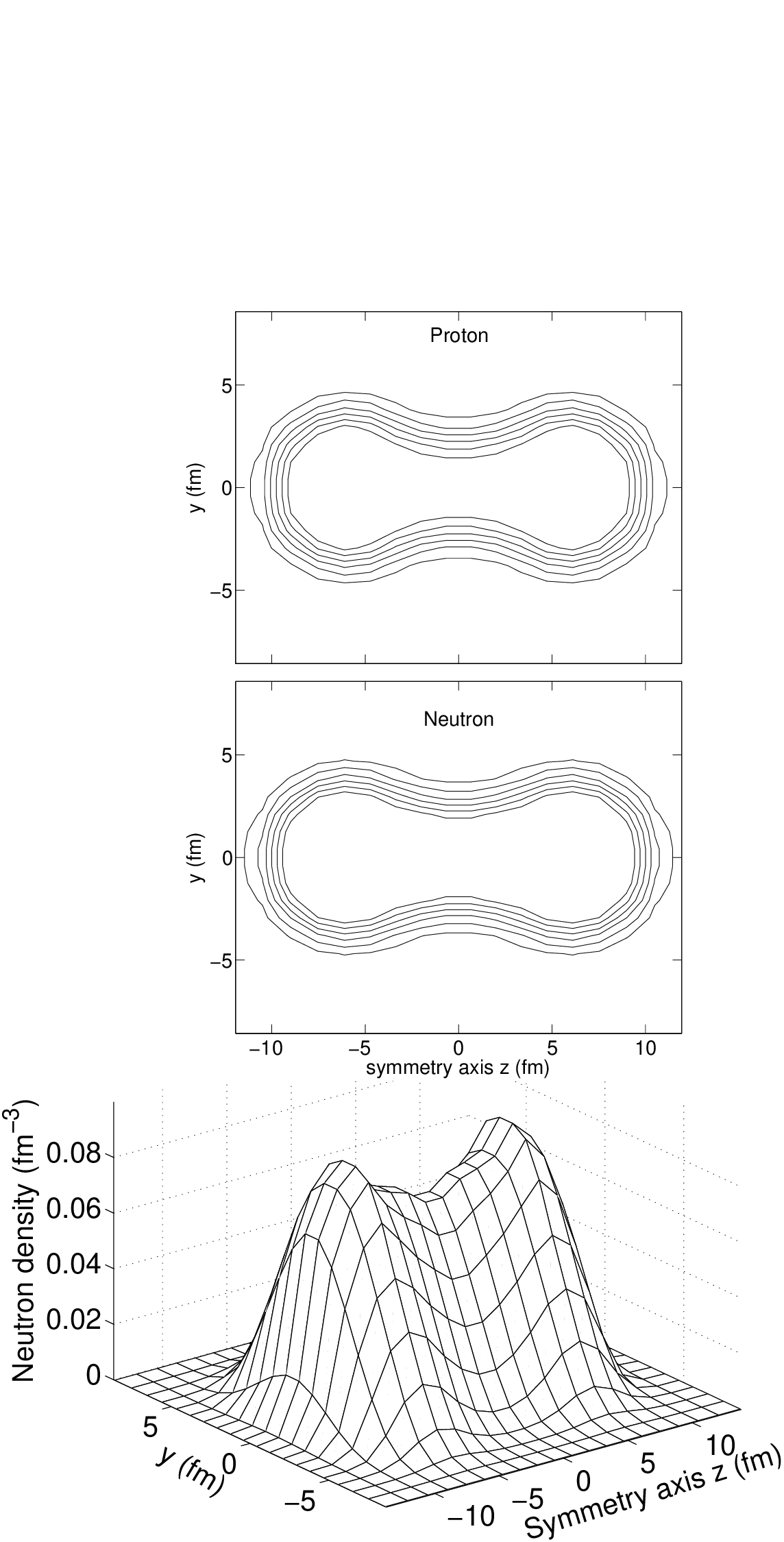}
\vspace{0.4cm}
\caption{The same as in Fig.\ \ref{density-Xe124}, but for yrast megadeformed state 
in $^{102}$Pd at rotational frequency $\Omega_x=0.95$ 
MeV. Two top panels show 2-dimensional plots of the proton and neutron density
distribution.}
\label{density-Pd102}
\end{figure}
%%%%%%%%%%%%%%%%%%%%%%%%%%%%%%%%%%%%%%%%%%%%%%%%%%%%%%%%%%%%%%%%%%%%%%%%%%%%%%%%%%

%%%%%%%%%%%%%%%%%%%%%%%%%%%%%%%%%%%%%%%%%%%%%%%%%%%%%%%%%%%%%%%%%%%%%%%%%%%%%%%%%%%
\begin{figure}[ht]
\includegraphics[angle=0,width=8.0cm]{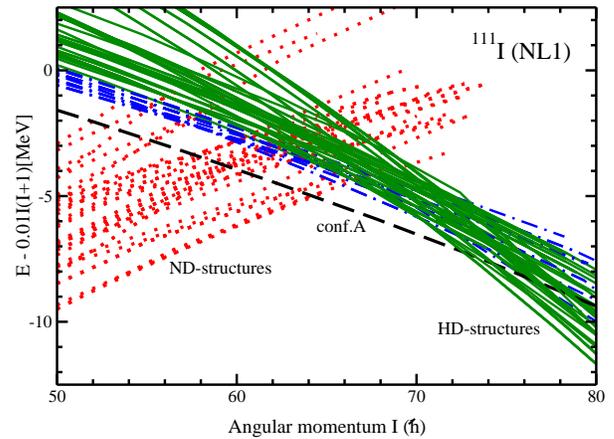}
\vspace{0.2cm}
\caption{(Color online) Energies of the calculated configurations relative to a smooth 
liquid drop reference $AI(I+1)$, with the inertia parameter $A=0.01$. Normal 
deformed (ND), SD and HD configurations are shown by dotted, dot-dashed and 
solid lines, respectively. Configuration A is shown by long-dashed line.}
\label{I111-eld-NL1}
\end{figure}
%%%%%%%%%%%%%%%%%%%%%%%%%%%%%%%%%%%%%%%%%%%%%%%%%%%%%%%%%%%%%%%%%%%%%%%%%%%%%%%%%%%

 The analysis of the Nilsson diagrams in Fig.\ \ref{108Cd-nilsson} already reveals some 
HD gaps in the single-particle spectra. At the values of $Q_0 \sim 17-20$ $e$b typical for the 
HD configurations in Cd isotopes (Fig.\ \ref{Icross-Cd}b), there are very large 
proton $Z=48$ and neutron $N=48$ HD shell gaps and smaller neutron gaps at $N=58$ and 60. 
In general, this figure suggests that the hyperdeformation will be more favoured in the 
nuclei with a similar number  of protons and neutrons because the proton and neutron
shell effects for the HD shapes will act coherently; this 
trend has already been seen in the crossings spins $I_{cr}^{HD}$ for different isotope 
chains in Sect.\  \ref{sys-Icros}.

The size of these gaps and their presence will be altered (especially, for medium and small size 
energy gaps) when the rotation and the self-consistent readjustment of the neutron and proton 
densities with the change of particle number are taken into account. Indeed, this is
seen in Fig.\ \ref{shell-gaps} which shows the energy gaps between the last occupied and first 
unoccupied single-particle orbitals as a function of the neutron number for different isotope 
chains.  The largest proton gap at $Z=48$ is seen in  Cd isotopes; its size is around 1.5 
MeV in proton-rich nuclei and it increases up to 3 MeV with the increase of neutron number. 
In other isotope chains, the size of the proton energy gap is smaller than in  Cd isotopes 
and it fluctuates around 1 MeV. For the majority of the nuclei, the size of the neutron energy gap 
fluctuates around 1 MeV. However, its size  increases up to 1.5 MeV in some nuclei and in 
$^{96}$Cd it reaches 2 MeV (see Fig.\ \ref{shell-gaps} for details).

   Taking into account that the proton and neutron HD shell gaps in $^{124}$Xe are around
1 MeV (Fig.\ \ref{Xe124-routh}) and considering the results for the density of the HD 
states in this nucleus as a reference (Sect.\ \ref{Xe124-neigh}), one can conclude that 
the analysis of the energy gaps suggests that in most of the nuclei the density of the HD bands 
will be high. For these nuclei, the observation of discrete HD bands using existing 
facilities is most likely not possible.  The only exceptions are Cd nuclei
and a few nuclei in which the size of at least one gap reaches 1.5 MeV (see Fig.\ 
\ref{shell-gaps} for details). For example, in  Cd nuclei the large size  of the $Z=48$ HD 
shell gap (especially, for nuclei in the valley of the $\beta$-stability) will make proton 
particle-hole excitations energetically expensive. As a consequence, the density of the HD 
bands has to be lower in Cd isotopes as compared with the one in other isotopes.

%%%%%%%%%%%%%%%%%%%%%%%%%%%%%%%%%%%%%%%%%%%%%%%%%%%%%%%%%%%%%%%%%%%%%%%%%%%%%%%%%%
\begin{figure}[h]
\includegraphics[angle=0,width=6.5cm]{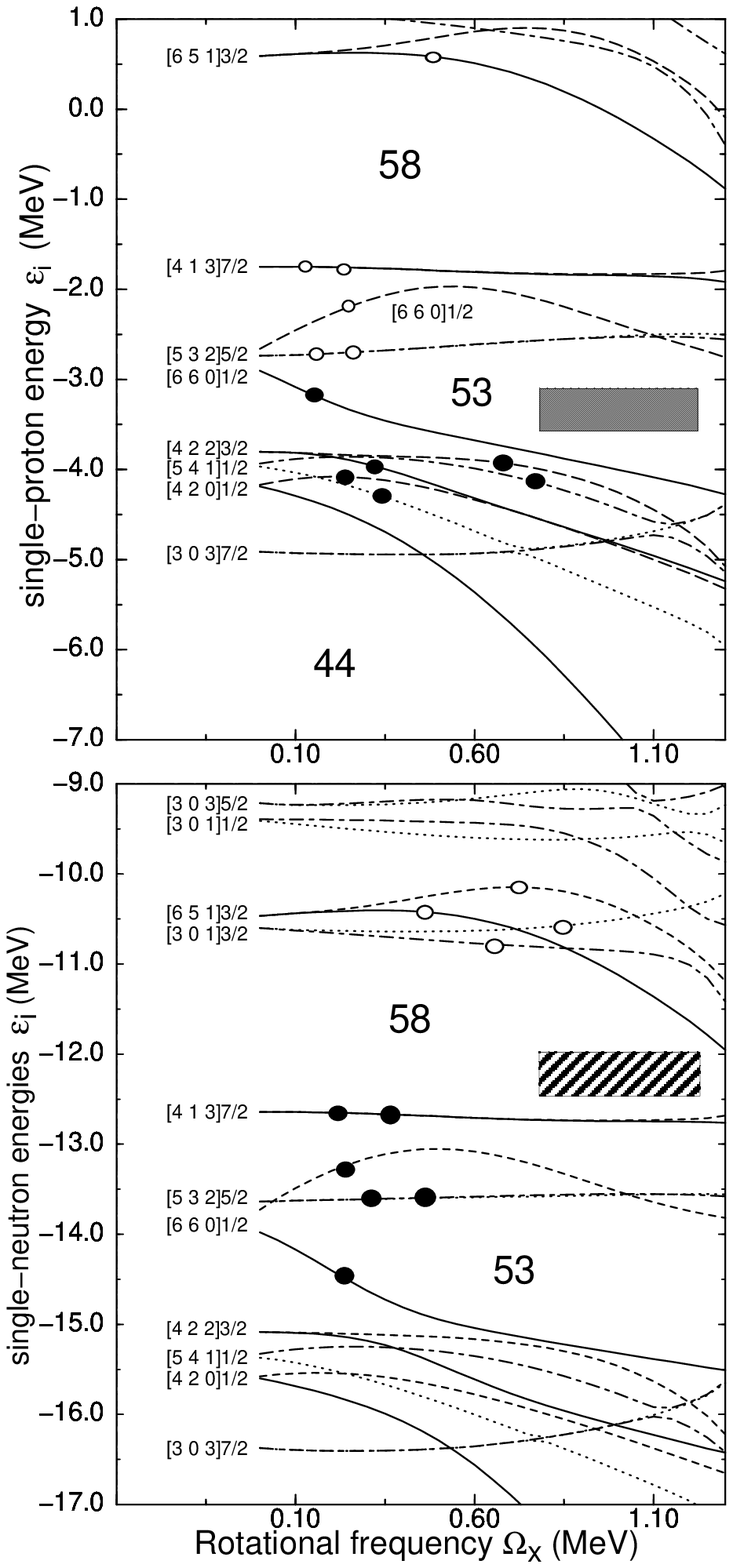}
\caption{ The same as in Fig.\ \ref{Xe124-routh}, but for the configuration A 
in $^{111}$I. Solid (open) circles indicate the orbitals occupied (emptied). 
The dashed box indicates the frequency range corresponding to the spin range 
$I=50 - 75\hbar$ in this configuration.}
\label{I111-routh}
\end{figure}
%%%%%%%%%%%%%%%%%%%%%%%%%%%%%%%%%%%%%%%%%%%%%%%%%%%%%%%%%%%%%%%%%%%%%%%%%%%%%%%%%%

  One has to remember that the high density of the HD bands is not necessarily a negative
factor. It favors the observation of the rotational patterns in the form of ridge-structures 
in three-dimensional  rotational mapped spectra as it has been seen in the HLHD experiment 
for a few nuclei \cite{Hetal.06}. The observation of ridge-structures as a function of proton
and neutron number, which seems to be feasible with existing experimental facilities such as 
GAMMASPHERE,  will provide invaluable information about HD at high spin.

  The importance of the necking degree of freedom for the high-spin HD states 
has been  studied in the MM approach in Refs.\ \cite{A180-HD-Chassman,C.01}. However, 
this degree of freedom has not been investigated in detail at high spin in self-consistent 
approaches so far. In order to fill 
this gap in our knowledge, the systematics of the self-consistent proton density 
distributions in the HD states obtained in the CRMF calculations are shown in Fig.\ 
\ref{density-sys}. One can see that in some nuclei such as $^{124}$Te, $^{130}$Xe, 
$^{132}$Ba the necking degree of freedom plays an important role, while others 
(for example, $^{100}$Mo and $^{136}$Ce) show no necking. The neck is typically less 
pronounced in the HD states of the lighter nuclei because of their smaller deformation 
(see also Fig.\ 5 in Ref.\ \cite{AF.05-108Cd}).  It becomes even more important in 
extremely deformed structures which according to the language of Ref.\ \cite{Dudek} can 
be described as megadeformed. Fig.\ \ref{density-Pd102} shows an example of density 
distribution for the megadeformed state in $^{102}$Pd, which becomes yrast at 
$I\sim 85\hbar$ in the CRMF calculations.  The neck is more pronounced in the proton 
subsystem  than in the neutron one both in the HD and megadeformed structures  due to 
the Coulomb repulsion of the segments. This is illustrated in Fig.\ \ref{density-Pd102}.
Our self-consistent calculations indicate that the shell  structure is also playing a role
in a formation of neck. For example, the neck is visible in $^{132}$Ba but is not 
seen in $^{116}$Ba (Fig.\ \ref{density-sys}). This is contrary to the fact that the 
calculated transition quadrupole moments of the HD states in these nuclei (Fig.\ 
\ref{Icross-Ce-Ba-Xe}d) and their density elongations (Fig.\ \ref{density-sys}) are 
comparable. These results indicate that, in general, the necking degree of freedom is 
important in the HD states and that it should be treated within the self-consistent 
approach which, in particular, allows different necking for the proton and neutron 
subsystems.

%%%%%%%%%%%%%%%%%%%%%%%%%%%%%%%%%%%%%%%%%%%%%%%%%%%%%%%%%%%%%%%%%%%%%%%%%%%%%%%%%%
\section{$^{111}$I nucleus: a candidate for a doubly magic extremely SD 
band.}
\label{I111-nucleus}
%%%%%%%%%%%%%%%%%%%%%%%%%%%%%%%%%%%%%%%%%%%%%%%%%%%%%%%%%%%%%%%%%%%%%%%%%%%%%%%%%%

%%%%%%%%%%%%%%%%%%%%%%%%%%%%%%%%%%%%%%%%%%%%%%%%%%%%%%%%%%%%%%%%%%%%%%%%%%%%%%%%%%
\begin{figure*}[ht]
\includegraphics[angle=0,width=13.0cm]{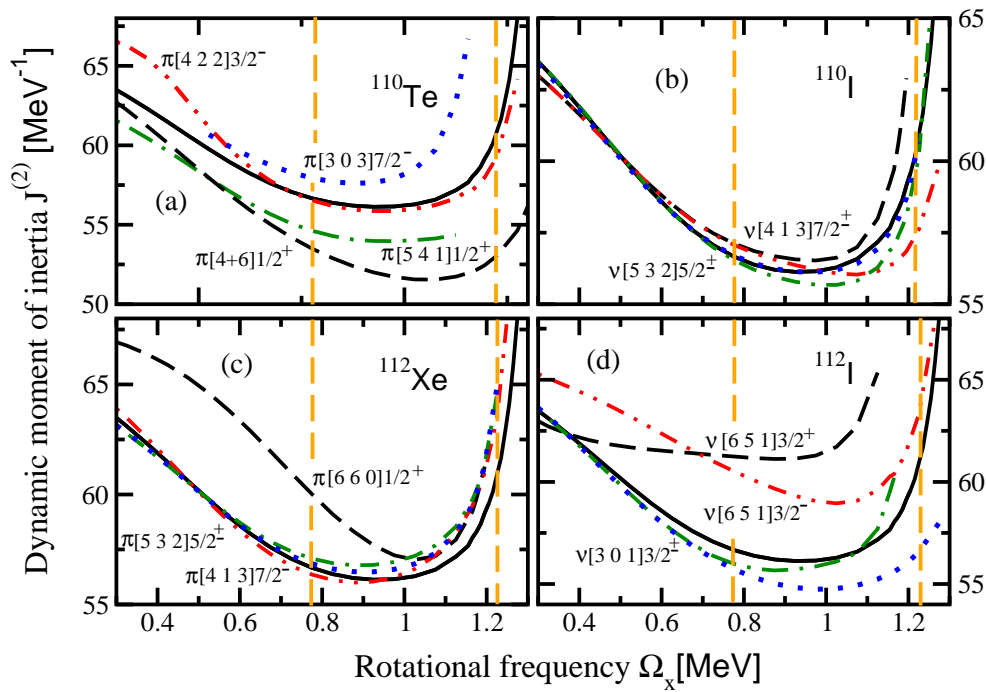}
\vspace{0.2cm}
\caption{\label{near-I111-j2} (Color online)
   The same as in Fig.\ \ref{near-Xe124-J2}, but for dynamic moments of inertia 
$J^{(2)}$ of the configurations used in  Fig.\ \ref{I111-ieff} below. Dynamic moments 
of inertia of the configuration A in $^{111}$I are shown by solid line in 
each panel. Vertical 
dashed lines indicate the frequency range corresponding to the spin range 
$I=50- 75\hbar$ in the configuration A of $^{111}$I. The $\pi [4+6]1/2$
label in panel (a) indicates the orbital with strong mixing of the $N=4$ and $N=6$ 
shells: this mixing predominantly emerges from the interaction of the $\pi [420]1/2$ 
and $\pi [660]1/2$ states.}
\end{figure*}
%%%%%%%%%%%%%%%%%%%%%%%%%%%%%%%%%%%%%%%%%%%%%%%%%%%%%%%%%%%%%%%%%%%%%%%%%%%%%%%%%%

   The results of the CRMF calculations for the configurations forming the 
yrast line or located close to it in energy are shown in Fig.\ \ref{I111-eld-NL1}.
According to the calculations, normal- and highly-deformed bands, many of which 
show the high triaxiality that is indicative of approaching band termination 
\cite{PhysRep}, dominate the yrast line up to $I \approx 64\hbar$. At higher spin, 
more deformed structures become yrast. The configuration A has the structure 
$\pi 6^1 \nu 6^2$ and is yrast in the spin range $I=64-73\hbar$: no hyperintruder 
$N=7$ orbitals are involved in its structure. In this spin range it is characterized 
by the transition quadrupole moment $Q_t \sim 15.7$ $e$b and by the $\gamma$-deformation 
of $\sim 1^{\circ}$. The normalized transition quadrupole moment in this system
is $Q^{norm}_t=11.7$ $e$b, thus, this band is approximately 35\% more deformed
than the SD band in $^{152}$Dy. As a consequence, in terms of deformation, this 
band can be characterized as an extremely superdeformed (ESD) band which is only 
slightly less deformed than the HD bands.

%%%%%%%%%%%%%%%%%%%%%%%%%%%%%%%%%%%%%%%%%%%%%%%%%%%%%%%%%%%%%%%%%%%%%%%%%%%
\begin{table}[h]
\caption{The size of the Z=53 and N=58 ESD shell gaps [in MeV] obtained 
with different parametrizations of the RMF Lagrangian for the configuration 
A in $^{111}I$ at spin $I=60\hbar$ (rotational frequency $\Omega_x \approx 0.96$ 
MeV).}
\newcommand{\m}{\hphantom{$-$}}
\newcommand{\cc}[1]{\multicolumn{1}{c}{#1}}
\renewcommand{\tabcolsep}{0.5pc} % enlarge column spacing
\renewcommand{\arraystretch}{1.4} % enlarge line spacing
\begin{tabular}{|c|c|c|c|c|} \hline
           & NL1       &   NL3    & NLZ    & NLSH \\ \hline   
Z=53       & 1.45      &  1.25    & 1.65   & 0.70 \\   
N=58       & 1.75      &  1.85    & 1.60   & 2.00 \\ \hline
\end{tabular}\\[2pt]
\label{Table-i111}
\end{table}
%%%%%%%%%%%%%%%%%%%%%%%%%%%%%%%%%%%%%%%%%%%%%%%%%%%%%%%%%%%%%%%%%%%%%%%%%

  In addition, the configuration A is well separated from the excited SD/HD configurations 
below $I\sim 73\hbar$ (see Fig.\ \ref{I111-eld-NL1}). This is due to the presence of 
the large $Z=53$ and $N=58$ ESD shell gaps in the single-particle spectra (see Fig.\ 
\ref{I111-routh}). In this configuration, all single-particle states below the $Z=53$ 
and $N=58$ ESD shell gaps are occupied by protons and neutrons, respectively. Thus, this 
ESD band is a doubly-magic one. This band appears as doubly-magic also in the 
calculations with widely used NL3 \cite{NL3} and NLZ \cite{NLZ} parametrizations of the 
RMF Lagrangian, see Table \ref{Table-i111}. Extensive calculations with the
NL3 parametrization (similar to the ones presented in Fig.\ \ref{I111-eld-NL1}) 
show  that this band become yrast at $I\sim 62\hbar$. The $Z=53$ ESD shell gap is 
smaller than 1 MeV only in the NLSH \cite{NLSH} parametrization of the RMF Lagrangian 
(see Table \ref{Table-i111}). However, it is known that the single-particle energies 
are not well described in this parametrization \cite{A250}. One should note, however, 
that the size of the ESD gaps in the configuration A of $^{111}$I  is somewhat smaller 
than the one for the yrast SD band in $^{152}$Dy (compare Fig.\ \ref{I111-eld-NL1} in 
the present manuscript with Fig.\ 3 in Ref.\ \cite{A150}; see also Figs.\ 4, 11, 12 
in Ref.\ \cite{ALR.98} obtained with different parametrizations of the RMF Lagrangian 
and relevant for $^{151}$Tb).

\begin{figure*}[ht]
\includegraphics[angle=0,width=13.0cm]{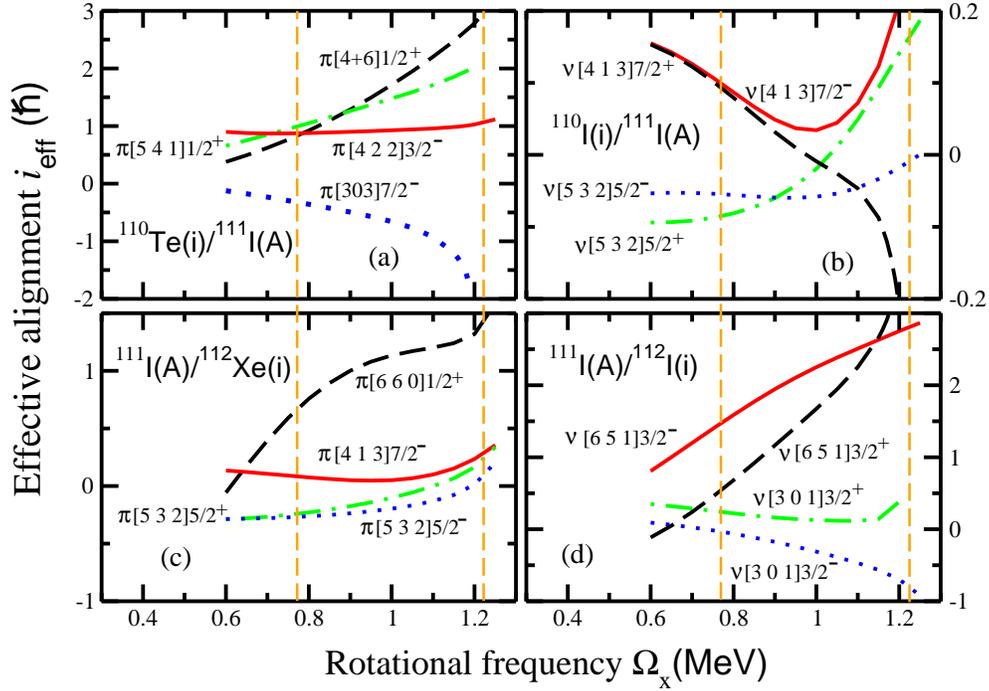}
\vspace{0.2cm}
\caption{(Color online) The same as in Fig.\ \ref{Xe124-ief}, but for effective alignments
of the single-particle orbitals in the vicinity of the $Z=53$ and $N=58$ SD shell 
gaps (see Fig.\ \ref{I111-routh}). The effective alignments are defined with respect
to the configuration A in $^{111}$I. Vertical dashed lines indicate the frequency 
range corresponding to the spin range $I=50 - 75\hbar$ in the configuration A of 
$^{111}$I.}
\label{I111-ieff}
\end{figure*}

  The dynamic moments of inertia of the configuration A in $^{111}$I and the 
configurations in neighboring nuclei are shown in Fig.\ \ref{near-I111-j2}. The 
increase of $J^{(2)}$ at $\Omega_x \sim 1.2$ MeV is in part due to unpaired 
band crossing caused by the interaction of the occupied $\nu [413]7/2^-$ and unoccupied 
$\nu [651]3/2^-$ orbitals (Fig.\ \ref{I111-routh}). A centrifugal stretching may also
contribute to this increase of $J^{(2)}$. The effect of the occupation of 
a single proton (neutron) intruder orbital
on the properties of the ESD bands is much more 
pronounced than that in the HD bands of the nuclei around 
$^{124}$Xe (see  Sect.\ \ref{Xe124-single}); the changes induced into dynamic 
moment of inertia reach at least  10\% of its absolute value for the 
$\pi [660]1/2^+$ (Fig.\ \ref{near-I111-j2}c), $\pi [4+6]1/2^+$ (Fig.\ \ref{near-I111-j2}a), 
$\nu [651]3/2^+$ (Fig.\ \ref{near-I111-j2}d) and $\nu [651]3/2^-$ (Fig.\ \ref{near-I111-j2}d)
orbitals. In a similar way, the effective alignments of these orbitals
as well as of the $\pi [541]1/2^+$ orbital show appreciable variations as a function 
of rotational frequency  (see Fig.\ \ref{I111-ieff}), reaching at least $1\hbar$ in 
the spin range of interest. This suggests that the  configuration assignment based 
on the effective alignment method will be more reliable in the case of ESD bands as 
compared with the HD bands in the nuclei around $^{124}$Xe (see Sect.\ \ref{Xe124-single} 
for a discussion of these methods). Relative properties of the dynamic moments  of inertia 
of two compared bands will also play a complimentary  role in the configuration assignment.

%%%%%%%%%%%%%%%%%%%%%%%%%%%%%%%%%%%%%%%%%%%%%%%%%%%%%%%%%%%%%%%%%%%%%%%%%%%%%%%%%%%
%%%%%%%%%%%%%%%%%%%%%%%%%%%%%%%%%%%%%%%%%%%%%%%%%%%%%%%%%%%%%%%%%%%%%%%%%%%%%%%%%%%

%%%%%%%%%%%%%%%%%%%%%%%%%%%%%%%%%%%%%%%%%%%%%%%%%%%%%%%%%%%%%%%%%%%%%%
\section{Conclusions}
\label{Concl-sect}
%%%%%%%%%%%%%%%%%%%%%%%%%%%%%%%%%%%%%%%%%%%%%%%%%%%%%%%%%%%%%%%%%%%%%%

  For the first time, the hyperdeformation at high spin has been studied
in a systematic way within the framework of a fully self-consistent theory:
the cranking relativistic mean field theory. The study covers even-even
nuclei in the $Z=40-58$ part of nuclear chart. The main results can be 
summarized as follows:

\begin{itemize}

\item
 The crossing spins $I_{cr}^{HD}$, at which the HD configurations become yrast, 
are lower for proton-rich nuclei. This is a feature seen in the 
most of studied isotope chains; by going from the $\beta$-stability valley 
towards the proton-drip line one can lower $I_{cr}^{HD}$ by approximately 
$10\hbar$. 

\item
   The density of the HD bands in the spin range where they are 
yrast or close to yrast is high in the majority of the cases. For such densities, 
the feeding intensity of an individual HD band will most likely drop below the 
observational limit of modern experimental facilities. This fact has to be taken 
into account when planning the experiments for a search of discrete HD bands. Our
calculations indicate Cd isotopes and few other nuclei with large shell gaps (see Sect.
\ref{gen-obs} for details) as the best candidates for a search of discrete HD bands.
An alternative candidate is the doubly magic extremely superdeformed band
in $^{111}$I, the deformation of which is only slightly lower than that of the 
HD bands, and which may be observed with existing experimental facilities.

\item
The high density of the HD bands will most likely favor the 
observation of the rotational patterns in the form of ridge-structures in 
three-dimensional rotational mapped spectra. The study of these patterns as
a function of proton and neutron numbers, which seems to be possible with 
existing facilities, will provide a valuable information about hyperdeformation 
at high spin.

\item
 With a very few exceptions, the HD shapes undergo a centrifugal stretching that 
results in an increase of the values of the transition quadrupole $Q_t$ and mass 
hexadecapole $Q_{40}$ moments as well as the dynamic moments of inertia $J^{(2)}$ 
with increasing rotational frequency. The kinematic moments of inertia $J^{(1)}$
show very small variations in the frequency range of interest. These are general 
features of the HD bands which distinguish them from the normal- and superdeformed 
bands. Such features have not been seen before in the calculations without pairing. 
In unpaired regime, the $Q_t$, $J^{(2)}$ and $J^{(1)}$ values decrease with rotational 
frequency in the SD configurations; the only exceptions are the regions of unpaired 
bands crossings.

\item 
 The individual properties of the single-particle orbitals are not lost at HD. In the 
future, they will allow the assignment of the configurations to the HD bands using the 
relative properties of different bands. Such methods of configuration assignment were 
originally developed for superdeformation. In contrast to the case of SD, our analysis 
in the $A\sim 125$ mass region shows that only simultaneous application of the methods 
based on effective alignments and relative transition quadrupole moments by comparing 
experimental and theoretical $(i_{eff},\Delta Q_t)$ values will lead to a reliable 
configuration assignment for the HD bands. Moreover, additional information on the 
structure of the HD bands will be obtained from the band crossing features; the cases of 
strong interaction of the bands in unpaired regime at HD will be more common as compared 
with the situation at SD.

\end{itemize}

 The physics of hyperdeformation at high spin is also defined by the fission barriers; 
the competition with fission certainly makes the population of the HD states difficult.
It is an important issue, which, however, goes beyond the scope of the current manuscript.
It is likely that the fission barriers are small or non-existent at the spins around 
$80-90\hbar$ in some of the studied nuclei; the observation of the HD bands then will not 
be possible in these systems. This problem definitely deserves a deeper attention; the 
study of the fission barriers at high spin typical for HD within the framework of the 
cranked relativistic Hartree-Bogoliubov theory is in its initial stage and the results 
will be presented in a forthcoming  manuscript.

%%%%%%%%%%%%%%%%%%%%%%%%%%%%%%%%%%%%%%%%%
\section{Acknowledgements}
%%%%%%%%%%%%%%%%%%%%%%%%%%%%%%%%%%%%%%%%%

   The help of C.\ W.\ Jang and J.\ Begnaud in performing numerical calculations is 
highly appreciated. The work was supported by the U.S. Department of Energy under 
grant DE-FG02-07ER41459. Stimulating discussions with Robert Janssens are gratefully
acknowledged.

\end{document}